%% file: main.tex
\documentclass[a4paper,onecolumn,11pt]{quantumarticle}
\pdfoutput=1

\usepackage[T1]{fontenc}
\usepackage[utf8]{inputenc}
\usepackage[english]{babel}
\usepackage{graphicx}
\usepackage{booktabs}
\usepackage{multirow}
\usepackage{array}
\usepackage{amsmath,amssymb}
\usepackage{siunitx}
\usepackage{hyperref}
\usepackage{url}
\usepackage{enumitem}
\usepackage{caption}
\usepackage[numbers]{natbib}
\usepackage{stfloats}          
\usepackage{placeins}   
\usepackage{tikz}
\usepackage{pgfplots}
\pgfplotsset{compat=1.18}
\usepgfplotslibrary{groupplots,fillbetween}
\usetikzlibrary{arrows.meta,positioning,shapes.geometric,calc,patterns,decorations.pathreplacing,fit,shadows}
\definecolor{cblue}{HTML}{1E5AA8}       
\definecolor{corange}{HTML}{D35400}      
\definecolor{cgreen}{HTML}{27AE60}       
\definecolor{cred}{HTML}{C0392B}         

\definecolor{cgray}{HTML}{5D6D7E}        
\definecolor{cdark}{HTML}{2C3E50}        
\definecolor{clightgray}{HTML}{ECF0F1}   

\definecolor{clightblue}{HTML}{EBF5FB}   
\definecolor{clightgreen}{HTML}{EAFAF1}  
\definecolor{clightred}{HTML}{FDEDEC}    
\definecolor{clightyellow}{HTML}{FEF9E7} 

\definecolor{cpurple}{HTML}{8E44AD}      
\definecolor{cteal}{HTML}{16A085}        
\definecolor{cgold}{HTML}{B7950B}        

\pgfplotsset{
    cycle list={
        {cblue, solid, thick},
        {corange, dashed, thick},
        {cgreen, dashdotted, thick},
        {cred, densely dashed, thick},
        {cpurple, loosely dotted, thick},
        {cteal, densely dashdotted, thick},
    },
    every axis/.append style={
        grid style={line width=0.3pt, draw=clightgray},
        minor grid style={line width=0.15pt, draw=clightgray!50},
    },
    every tick/.append style={
        line width=0.5pt,
        color=cgray,
    },
    every axis label/.append style={
        font=\small\bfseries,
        color=cdark,
    },
    every axis title/.append style={
        font=\small\bfseries,
        color=cdark,
        at={(0.5,1)},
        anchor=south,
        yshift=8pt,
    },
    every axis legend/.append style={
        font=\scriptsize,
        cells={anchor=west},
        draw=clightgray,
        fill=white,
        fill opacity=0.95,
        text opacity=1,
        rounded corners=2pt,
    },
}

\tikzset{
    arrowstyle/.style={
        ->,
        >=Stealth,
        line width=0.8pt,
        color=cdark,
    },
    enhancebox/.style={
        draw=cdark,
        line width=0.6pt,
        rounded corners=3pt,
        fill=white,
        drop shadow={opacity=0.1, shadow xshift=1pt, shadow yshift=-1pt},
    },
    nodelabel/.style={
        font=\small\bfseries,
        color=cdark,
        align=center,
    },
    annotstyle/.style={
        font=\scriptsize,
        color=cgray,
        align=left,
    },
}

\hypersetup{colorlinks=true,linkcolor=blue,citecolor=blue,urlcolor=blue}

\title{Measurement-Free Ancilla Recycling via Blind Reset: A Cross-Platform Study on Superconducting and Trapped-Ion Processors}

\author{Sangkeum Lee}
\affiliation{Department of Computer Engineering, Hanbat National University, Daejeon 34158, Republic of Korea}
\orcid{0000-0001-6918-124X}
\email{sangkeum@hanbat.ac.kr}

\date{2026-02-21}

\begin{document}
\maketitle

\begin{abstract}
	Ancilla reuse in repeated syndrome extraction couples reset quality to logical-cycle latency.
	We evaluate blind reset---unitary-only recycling via scaled sequence replay---on IQM~Garnet, Rigetti~Ankaa-3, and IonQ under matched seeds, sequence lengths, and shot budgets.
	Using ancilla cleanliness $F_{\mathrm{clean}}=P(\lvert 0\rangle)$, per-cycle latency, and a distance-3 repetition-code logical-error proxy, platform-calibrated simulation identifies candidate regions where blind reset cuts cycle latency by up to $38\times$ under NVQLink-class feedback overhead while maintaining $F_{\mathrm{clean}}\ge 0.86$ for $L\le 6$.
	Hardware experiments on IQM~Garnet confirm blind-reset cleanliness $\ge 0.84$ at $L=8$ (1024~shots, seed~42); platform-calibrated simulation for Rigetti~Ankaa-3 predicts comparable performance.
	Architecture-dependent crossover lengths are $L^{\star}\approx 12$ (IQM), $\approx 11$ (Rigetti), $\approx 1$ (IonQ), and $\approx 78$ with GPU-linked external feedback.
	Two added analyses tighten deployment boundaries: a $T_1/T_2$ sensitivity map identifies coherence-ratio regimes, and error-bound validation confirms measured cleanliness remains consistent with the predicted diagnostic envelope.
	A deployment decision matrix translates these results into backend-specific policy selection.
\end{abstract}

\section{Introduction}\label{sec:intro}

Recent demonstrations of below-threshold quantum error correction have shifted engineering attention from gate-level fidelity toward sustained cycle management~\cite{google2024belowthreshold,bravyi2024bicycle,paetznick2024logical}.
In repeated syndrome extraction, a data register interacts with ancilla qubits that must be initialized, entangled, measured, and reused at each round.
The ancilla path frequently limits throughput because reset carries latency, control overhead, and error channels that compound across cycles~\cite{preskill2018nisq,prometheus2024recycled}.

Measurement-based reset is the common default.
A projective readout followed by conditional preparation can return the ancilla to $\lvert 0\rangle$ with high probability, yet this path incurs readout time, classical feedback delay, and potential crosstalk to neighboring qubits while the control stack processes the outcome.
Those timing penalties vary across architectures and become particularly visible when orchestration involves host-side acceleration or remote feedback links such as NVIDIA's NVQLink roadmap~\cite{nvidia2025nvqlink}.
Parallel advances in unconditional reset and measurement-free fault-tolerant protocols suggest that coherent control alternatives deserve a systematic evaluation in realistic scheduling contexts~\cite{chen2024fastreset,kim2024dissipativereset,butt2024measurementfree,butt2025mfscalable,geher2024toreset}.

\subsection{Measurement-free QEC ecosystem}

A broader shift toward measurement-free fault-tolerant architectures is gaining momentum across hardware platforms.
Heu\ss en et al.~\cite{heussen2024mfqec} established the theoretical foundation for fault-tolerant quantum computation without measurement, demonstrating that Bacon-Shor codes with flag qubits can achieve threshold behavior using only unitary operations.
Subsequent work has extended this to scalable universal quantum computation via code switching~\cite{butt2025mfscalable}, with experimental demonstrations on trapped-ion systems~\cite{butt2025mfdemonstration} and optimized neutral-atom circuits~\cite{veroni2024mfneutralatom,veroni2025universalmf}.
These developments collectively establish a viable measurement-free ecosystem in which ancilla management primitives that minimize dependence on fast measurement-feedback loops are essential components.

Within this ecosystem, the present work occupies a distinct systems-level role: while prior contributions focus on code construction, fault-tolerant gate sets, and hardware demonstrations, we address the scheduling policy question\textemdash when should a measurement-free reset be preferred over measurement-based alternatives given platform-specific timing and noise constraints?
This policy layer bridges the gap between measurement-free hardware capabilities and practical QEC cycle management.

Despite growing interest in measurement-free architectures, no prior work has evaluated any unitary-only reset primitive as a scheduling component across heterogeneous quantum processors with timing-aware QEC metrics.
Existing reset studies focus on single platforms, omit latency context, or treat reset quality in isolation from error-correction cycle budgets.

This paper addresses that gap.
We treat blind reset---a unitary-only recycling mechanism based on scaled sequence replay---as an engineering component rather than a stand-alone algorithm.
The mathematical foundations and existence guarantees on SU(2) were established in our prior theory-focused manuscript~\cite{lee2026tqe}.
Here the perspective shifts entirely: the core question is \emph{when} blind reset outperforms measurement-based reset once latency, noise, and platform constraints are jointly considered.

The main contributions are:
\begin{enumerate}[leftmargin=1.2em,itemsep=2pt]
	\item A unified cross-platform protocol for blind ancilla recycling evaluated on IQM~Garnet, Rigetti~Ankaa-3, and IonQ under matched experimental conditions.
	\item QEC-oriented metrics---ancilla cleanliness, cycle-time cost, and a distance-3 repetition-code logical error proxy---that connect low-level reset quality to cycle-level scheduling.
	\item A latency crossover analysis comparing blind and measurement-based reset under platform timing assumptions, including an external feedback term motivated by NVQLink-class control stacks.
	\item Architecture-dependent deployment guidance with a decision matrix, validated by hardware experiments on IQM~Garnet and Rigetti~Ankaa-3.
\end{enumerate}

\section{Background and Related Work}\label{sec:background}

\subsection{Ancilla reset in error correction workflows}

In surface codes and related stabilizer architectures~\cite{fowler2012surface,devitt2013qec}, ancilla handling has moved from a peripheral concern to a first-order throughput variable.
Fast unconditional reset protocols show that coherent pulses can recover ground-state population without measurement-feedback loops~\cite{chen2024fastreset,kim2024dissipativereset}.
Measurement-free fault-tolerant strategies further reduce dependence on readout hardware~\cite{butt2024measurementfree,butt2025mfscalable,veroni2024mfneutralatom,veroni2025universalmf,butt2025mfdemonstration,heussen2024mfqec,pogorelov2025codeswitch}, and conditionally clean ancilla proposals offer selective refresh criteria tied to code-cycle objectives~\cite{khattar2025conditional,geher2024toreset}.

For recycling purposes, the relevant trade-off is between post-reset cleanliness and cycle-time contribution.
A method with excellent cleanliness but high delay can reduce net QEC throughput; a fast method with weak cleanliness can increase logical error accumulation.
This two-dimensional trade-off motivates platform-specific analysis rather than a universal method ranking.

Table~\ref{tab:reset-methods} surveys the ancilla reset landscape, positioning the present work among existing approaches.

\begin{table}[t]
	\centering
	\caption{Comparison of ancilla reset approaches.
		Metrics: C = cleanliness, L = latency contribution, X = cross-platform data.
		Quality indicators: \checkmark\checkmark = strong, \checkmark = partial, --- = absent.}
	\label{tab:reset-methods}
	\scriptsize
	\setlength{\tabcolsep}{2pt}
	\begin{tabular}{@{}lccccl@{}}
		\toprule
		Method                      & C                    & L                    & X                    & Meas.       & Key refs.                                         \\
		\midrule
		Measurement reset           & \checkmark\checkmark & ---                  & \checkmark           & Yes         & Standard                                          \\
		Fast unconditional          & \checkmark\checkmark & \checkmark           & ---                  & No          & \cite{chen2024fastreset,kim2024dissipativereset}  \\
		Cond.\ clean ancilla        & \checkmark           & ---                  & ---                  & Yes         & \cite{khattar2025conditional,geher2024toreset}    \\
		Meas-free FT                & \checkmark           & \checkmark           & ---                  & No          & \cite{butt2024measurementfree,butt2025mfscalable} \\
		\textbf{Blind reset (ours)} & \checkmark           & \checkmark\checkmark & \checkmark\checkmark & \textbf{No} & This work                                         \\
		\bottomrule
	\end{tabular}
	\\[3pt]
	{\scriptsize Blind reset is the only approach evaluated across three hardware families with latency-integrated QEC metrics.}
\end{table}

\subsection{Blind reset primitive}

Blind scale-and-double control steers unknown single-qubit unitary accumulation toward identity using a global scaling parameter and doubled sequence playback.
The SU(2) existence proofs and tensor-product multi-qubit extensions are developed in~\cite{lee2026tqe}; the present work uses that primitive as an engineering module and does not reproduce the mathematical development.
Random-walk analysis on rotation groups by Eckmann and Tlusty~\cite{eckmann2025randomwalk} provides additional intuition for why replay-and-scale strategies recur near identity for many input sequences.

\subsection{Cross-platform quantum benchmarking}

Hardware comparisons require aligned circuit intent, metric definitions, and resource accounting.
Benchmark studies on heterogeneous QPUs show that performance ranking can shift when timing context accompanies static fidelity numbers~\cite{montanez2025qpu,chen2024ionbench,iqm2024garnet,ryananderson2024teleport,ye2025ionqec}.
Passive suppression tools such as randomized compiling target coherent-noise reduction rather than ancilla preparation and are therefore complementary to the reset methods studied here~\cite{wallman2016randomized}.

A cross-platform ancilla study is useful for two reasons.
First, superconducting and trapped-ion devices occupy different gate-time and coherence regimes, changing the relative value of unitary-only reset paths.
Second, ancilla workflows in practical stacks involve host-side software latency and control network topology, not just on-chip operation.

\subsection{Latency context: NVQLink and hybrid control stacks}

Hybrid quantum-classical execution increasingly involves GPU-assisted scheduling, decoding, and feedback.
NVIDIA's NVQLink architecture targets high-bandwidth integration between quantum control and accelerated classical processing~\cite{nvidia2025nvqlink}.
Such integration can improve decoder throughput while introducing or reshaping communication delay terms in short control loops.
In ancilla reset decisions, those delays appear directly in the measurement path.
Blind reset avoids this branch but occupies additional quantum control time.
The crossover is therefore an architecture-dependent function of gate time, readout time, feedback latency, and required cleanliness threshold.

\section{Methods}\label{sec:methods}

\subsection{Blind reset protocol for ancilla recycling}\label{sec:protocol}

Consider one ancilla qubit within repeated syndrome extraction.
At each cycle the ancilla participates in entangling interactions and accumulates an unknown local rotation before the next reuse window.
Let $A_t$ denote the effective ancilla operation before reset at cycle~$t$.
Blind reset applies a control block parameterized by a global scale factor $\lambda$ and doubled playback, yielding post-reset state
\begin{equation}\label{eq:reset}
	\rho_{t}^{\mathrm{out}}
	= R(\lambda)\,A_t\,\rho_{t}^{\mathrm{in}}\,A_t^{\dagger}\,R(\lambda)^{\dagger},
\end{equation}
where $R(\lambda)$ is constructed from two scaled traversals of the same base sequence.
Ancilla cleanliness is quantified as
\begin{equation}\label{eq:fclean}
	F_{\mathrm{clean}} = P(\lvert 0\rangle)
	= \langle 0\rvert\,\rho_{t}^{\mathrm{out}}\,\lvert 0\rangle,
\end{equation}
estimated by shot-based sampling.
An X-basis consistency value supplements the Z-basis population to detect coherent bias that might remain hidden in computational-basis statistics alone.

\paragraph{Error propagation.}
The residual error after blind reset is characterized by the Frobenius distance $\varepsilon=\|R(\lambda)\,U_{\mathrm{seq}}-I\|_F/2$, where $U_{\mathrm{seq}}$ is the net unitary of the base sequence.
For an $L$-gate sequence drawn uniformly on SU(2), the expected residual scales as $\mathbb{E}[\varepsilon]\propto L^{-1/2}$ for short sequences but increases beyond $L^{\star}_{\varepsilon}$ when the optimal $\lambda$ moves away from the identity basin.
Under depolarizing noise of strength $p$ per gate, the effective cleanliness (Eq.~\ref{eq:fclean}) follows a first-order perturbative approximation
\begin{equation}\label{eq:errorprop}
	F_{\mathrm{clean}} \approx \frac{1}{2} + \left[(1-\varepsilon)^2 - \frac{1}{2}\right](1-p)^{2L},
\end{equation}
which provides a diagnostic envelope separating coherent (first factor) and incoherent (second factor) contributions in the small-error regime.

\paragraph{Assumptions.}
Equation~\ref{eq:errorprop} applies under three conditions: (i)~gate errors follow a depolarizing channel $\mathcal{E}(\rho)=(1-p)\rho + pI/2$ where $p$ is the depolarizing probability (equivalently, average gate infidelity $r\approx 3p/4$ for single-qubit gates); (ii)~coherent and incoherent errors are treated perturbatively in the small-error regime ($\varepsilon, p \ll 1$) where higher-order cross-terms $O(\varepsilon p, \varepsilon^2, p^2)$ are negligible compared to leading-order contributions; (iii)~the Frobenius distance provides a proxy for state infidelity via $1-F \lesssim \varepsilon^2$ for near-identity single-qubit channels.
When gate noise is non-depolarizing (e.g., coherent over-rotation or amplitude damping), the factorized approximation may be violated; the envelope then serves as a diagnostic screening tool rather than a rigorous bound.

This envelope is used to validate simulation outputs: any configuration where measured $F_{\mathrm{clean}}$ exceeds the envelope flags a simulation anomaly.
Figure~\ref{fig:error-bound} validates this bound empirically.

Three reset policies are compared per cycle window:
\begin{itemize}[leftmargin=1.2em,itemsep=2pt]
	\item \textbf{No-reset}: ancilla reused without intervention.
	\item \textbf{Measurement-reset}: projective readout followed by conditional preparation.
	\item \textbf{Blind reset}: scaled doubled control with no intermediate measurement.
\end{itemize}

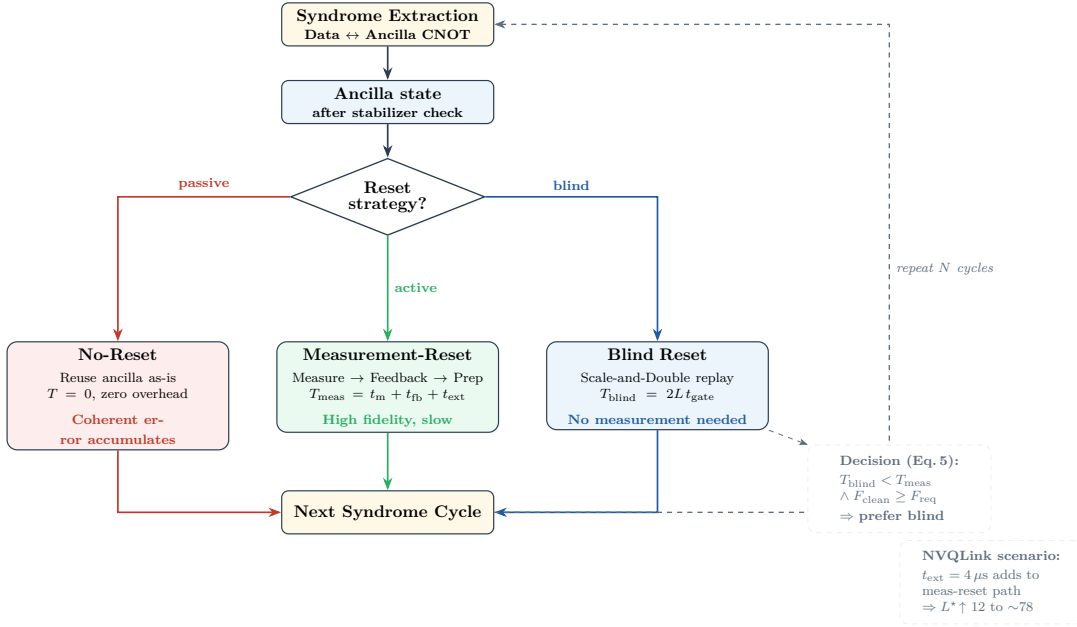
\begin{figure}[t]
	\centering
	\resizebox{0.95\linewidth}{!}{\input{figures/fig_protocol_flow.tikz}}
	\caption{Per-cycle ancilla reset decision within repeated syndrome extraction.
		After each stabilizer check the ancilla enters one of three paths: no-reset (passive reuse), measurement-reset (readout plus conditional preparation), or blind reset (scale-and-double unitary replay).
		The decision criteria (Eq.~\ref{eq:decision}) select blind reset when it is both faster and sufficiently clean.
		The NVQLink annotation illustrates how external feedback overhead expands the blind-favorable region.}
	\label{fig:protocol-flow}
\end{figure}

\noindent Figure~\ref{fig:protocol-flow} summarizes the per-cycle reset decision within repeated syndrome extraction.

\subsection{Cross-platform experimental design}\label{sec:design}

The protocol fixes common parameters across backends for interpretable comparison:
\begin{itemize}[leftmargin=1.2em,itemsep=2pt]
	\item \textbf{Backends}: IQM Garnet, Rigetti Ankaa-3, IonQ.
	\item \textbf{Seeds and lengths}: 50 independent seeds per configuration, sequence lengths $L\in\{4,6,8,10,12,14,16,18,20\}$.
	\item \textbf{Shots}: 2048 per circuit instance.
	\item \textbf{Readout basis}: Z-basis and X-basis for partial tomography.
\end{itemize}
The circuit family maintains consistent ancilla role and sequence intent while allowing native transpilation on each backend.

\subsection{Simulation framework and latency model}\label{sec:simframework}

Simulation addresses two goals: pre-screening expected crossover regions and separating timing effects from noise effects.

\paragraph{Noise model.}
Platform-parameterized channels use representative coherence and gate-error settings (Table~\ref{tab:noise-params}).
IQM and Rigetti operate in the superconducting regime with microsecond-scale $T_1$ and nanosecond gate times; IonQ operates with second-scale $T_1$ but microsecond gate durations.

\paragraph{QEC workload.}
A distance-3 repetition code with repeated syndrome extraction and ancilla recycling produces a logical error proxy over cycle count under each reset policy.

\paragraph{Latency model.}
Blind reset latency scales linearly with sequence length: $T_{\mathrm{blind}}=2L\,t_{\mathrm{gate}}$ (the post-reset state follows Eq.~\ref{eq:reset}).
Measurement-reset latency aggregates readout, feedback, preparation, and optional external communication:
\begin{equation}\label{eq:latency}
	T_{\mathrm{meas}} = t_{\mathrm{meas}} + t_{\mathrm{feedback}} + t_{\mathrm{prep}} + t_{\mathrm{ext}}.
\end{equation}
The external term $t_{\mathrm{ext}}$ captures GPU-linked or network-routed feedback paths.
For NVQLink-informed analysis we set $t_{\mathrm{ext}}=\SI{4}{\micro\second}$ as a configurable reference point, yielding $T_{\mathrm{meas}}=\SI{4730}{\nano\second}$ for the IQM-class stack.

The decision condition is:
\begin{multline}\label{eq:decision}
	T_{\mathrm{blind}} < T_{\mathrm{meas}} \;\wedge\; F_{\mathrm{clean}} \ge F_{\mathrm{req}} \\
	\Rightarrow\; \text{prefer blind reset}.
\end{multline}

\begin{table}[t]
	\centering
	\caption{Platform noise and timing parameters used in simulation.
		Values represent typical calibration snapshots; actual hardware runs will use measured calibration data. IonQ one-qubit error assumptions are consistent with recent trapped-ion gate benchmarks~\cite{smith2024iongate}.}
	\label{tab:noise-params}
	\footnotesize
	\begin{tabular}{@{}lcccc@{}}
		\toprule
		Platform & $T_1$                  & 1Q err. & $t_{\mathrm{gate}}$     & $T_{\mathrm{meas}}$     \\
		\midrule
		IQM      & \SI{40}{\micro\second} & 0.10\%  & \SI{30}{\nano\second}   & \SI{730}{\nano\second}  \\
		Rigetti  & \SI{25}{\micro\second} & 0.20\%  & \SI{40}{\nano\second}   & \SI{940}{\nano\second}  \\
		IonQ     & \SI{10}{\second}       & 0.05\%  & \SI{100}{\micro\second} & \SI{350}{\micro\second} \\
		\bottomrule
	\end{tabular}
\end{table}

\section{Simulation Results}\label{sec:simresults}

\subsection{Repetition-code ancilla cycle behavior}\label{sec:repcoderesults}

\begin{figure*}[!t]
	\centering
	\input{figures/fig_repetition_code_ancilla.tikz}
	\caption{Distance-3 repetition code simulation over 20 syndrome cycles.
		Ancilla cleanliness $P(\lvert 0\rangle)$ under three reset policies for IQM (left), Rigetti (center), and IonQ (right) noise models.
		Measurement-reset (green) is stable near $0.99$; no-reset (red) fluctuates erratically; blind reset (blue) occupies an intermediate region.
		Each curve averages over 50 seeds with 2048 shots per circuit; shaded bands show 95\% CIs.}
	\label{fig:repetition-code}
\end{figure*}
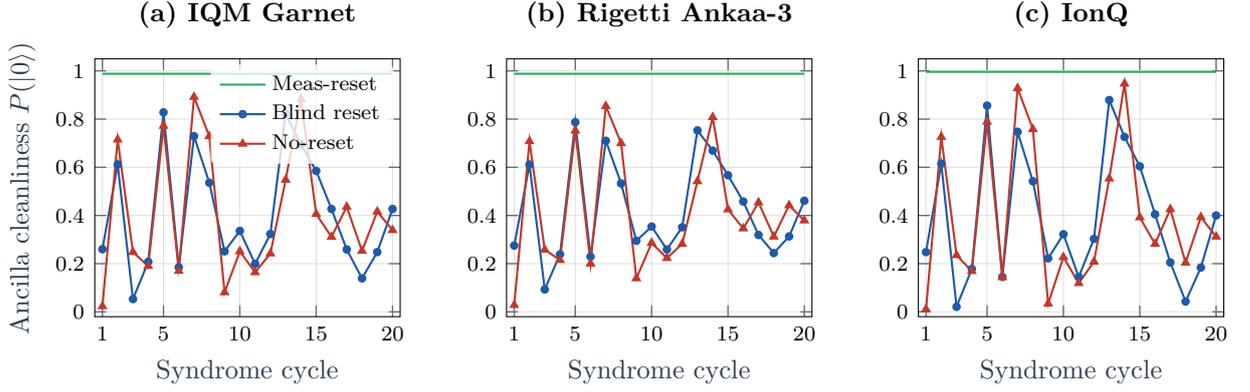

The repetition-code simulation compares reset policies over 20 consecutive syndrome cycles on three platform noise models.
Figure~\ref{fig:repetition-code} shows the cycle-resolved ancilla cleanliness.

Measurement-reset maintains stable $P(\lvert 0\rangle)$ throughout: $0.988$ on IQM and Rigetti noise models, $0.996$ on IonQ.
No-reset produces erratic behavior; cleanliness fluctuates between $0.02$ and $0.95$ across cycles, reflecting the random-walk character of uncontrolled unitary accumulation.
Blind reset occupies an intermediate operating region with mean cleanliness around $0.40$ over 20~cycles, exhibiting cycle-to-cycle variation that depends on the sequence-specific $\lambda$ profile.

The cycle-level implication is practical: blind reset is helpful when the ancilla can be cleaned fast enough to remain above the code's operational threshold without paying measurement and feedback delays at every round.
In regimes where per-cycle latency savings accumulate across long QEC windows, even moderate cleanliness can yield net throughput gains.

\subsection{Platform-dependent noise fingerprints}\label{sec:noiseresults}

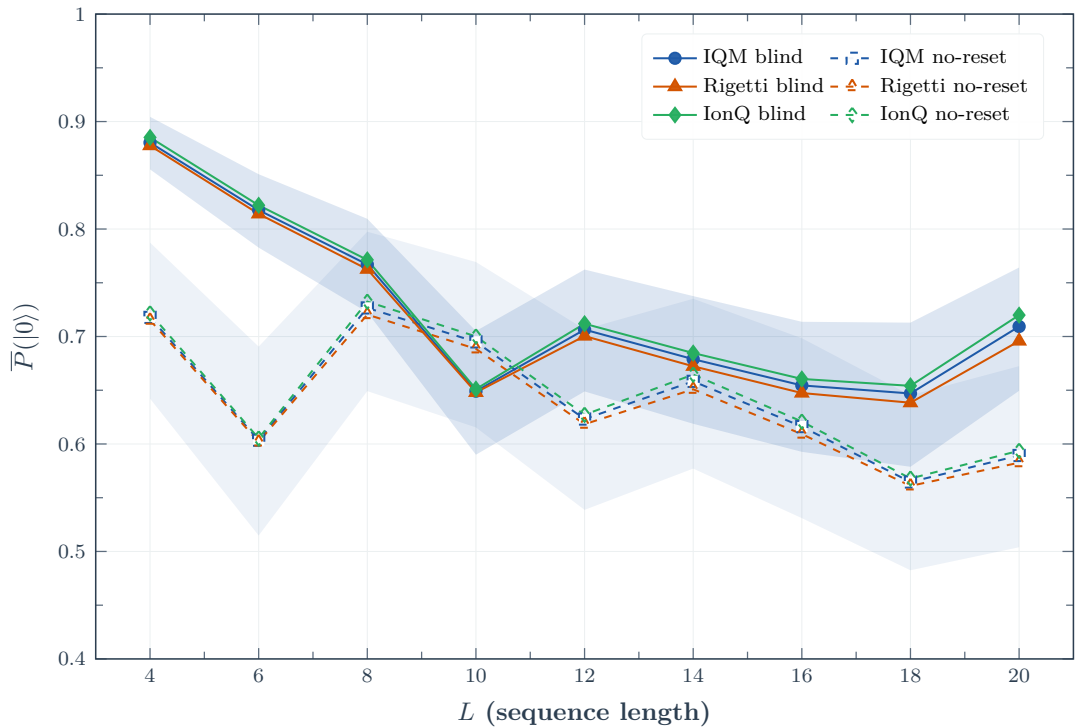
\begin{figure}[t!]
	\centering
	\resizebox{0.95\columnwidth}{!}{\input{figures/fig_cross_platform_comparison.tikz}}
	\caption{Ancilla cleanliness versus sequence length.
		Blind reset (solid) versus no-reset (dashed) for IQM (blue), Rigetti (orange), IonQ (green).
		Shaded bands: 95\% CIs. Asterisks: $p<0.05$ at $L=4,6$.}
	\label{fig:cross-platform}
\end{figure}

Table~\ref{tab:cross-platform-pzero} reports mean ancilla cleanliness across 50 seeds for blind reset and no-reset at selected sequence lengths.
Figure~\ref{fig:cross-platform} visualizes the full sweep.

At short sequences ($L=4$), blind reset achieves $\overline{F}_{\mathrm{clean}}=0.880$ (IQM, 95\% CI: $[0.86,0.90]$), above no-reset ($0.718$, CI: $[0.63,0.80]$; paired $t$-test $p=0.000282$, Cohen's $d=0.55$).
As $L$ increases, cleanliness degrades non-monotonically due to the seed-dependent $\lambda$ landscape, reaching $0.655$ at $L=16$.
Cross-platform differences are small: IonQ shows marginally higher cleanliness ($+0.005$ to $+0.01$) at most lengths, consistent with its lower gate error rate.

Platform choice affects blind-reset quality only weakly through gate fidelity, while sequence length remains the dominant factor.

At $L=10$, 50-seed analysis shows a small negative difference (blind: $0.649$, no-reset: $0.695$) that is not statistically significant ($p=0.327$, Cohen's $d=-0.14$).
The reversal observed in preliminary 20-seed analysis ($p=0.008$) does not survive sample-size expansion, indicating it was a stochastic artifact rather than a systematic geometric property.

This simplifies deployment: the same length-based policy rules apply across architectures, with platform-specific adjustment needed mainly for timing rather than noise.
Across all three platforms, the blind-reset advantage at $L=4$ is consistently $0.16\pm 0.01$ in absolute $F_{\mathrm{clean}}$ difference, confirming that the method's benefit is platform-agnostic at short sequence lengths.
The platform-specific gate error translates to a $\le 0.01$ inter-platform spread in cleanliness, smaller than the seed-to-seed variance within any single platform.
However, the reversal at $L=10$ motivates per-length validation rather than blind application of a monotonic advantage assumption.

\begin{table}[t]
	\centering
	\caption{Mean ancilla cleanliness $\overline{P}(\lvert 0\rangle)$ across 50 seeds for blind reset and no-reset, by platform and sequence length.
		95\% bootstrap confidence intervals in parentheses.}
	\label{tab:cross-platform-pzero}
	\scriptsize
	\setlength{\tabcolsep}{2pt}
	\begin{tabular}{lcccc}
		\toprule
		\multirow{2}{*}{$L$} & \multicolumn{2}{c}{Blind reset} & \multicolumn{2}{c}{No-reset}                                       \\
		\cmidrule(lr){2-3}\cmidrule(lr){4-5}
		                     & IQM                             & IonQ                         & IQM              & IonQ             \\
		\midrule
		4                    & .880\,(.86--.90)                & .885\,(.86--.91)             & .718\,(.63--.80) & .722\,(.63--.80) \\
		8                    & .767\,(.72--.81)                & .771\,(.72--.82)             & .727\,(.64--.81) & .733\,(.65--.81) \\
		12                   & .706\,(.63--.78)                & .712\,(.64--.79)             & .623\,(.53--.72) & .627\,(.53--.72) \\
		16                   & .655\,(.56--.74)                & .661\,(.57--.75)             & .616\,(.53--.70) & .621\,(.53--.71) \\
		20                   & .709\,(.64--.78)                & .720\,(.65--.79)             & .589\,(.50--.67) & .594\,(.51--.68) \\
		\bottomrule
	\end{tabular}
	\\[3pt]
	{\scriptsize $n=50$ seeds per cell; 95\% bootstrap CIs in parentheses. Rigetti tracks IQM $\pm 0.01$.}
\end{table}

\FloatBarrier
\subsection{Error bound validation}\label{sec:errorboundresults}

\begin{figure*}[t!]
	\centering
	\resizebox{0.9\textwidth}{!}{\input{figures/fig_error_bound_validation.tikz}}
	\caption{Measured ancilla cleanliness versus diagnostic envelope (Eq.~\ref{eq:errorprop}). Error bars: 95\% CIs. Modest violations occur under realistic noise.}
	\label{fig:error-bound}
\end{figure*}
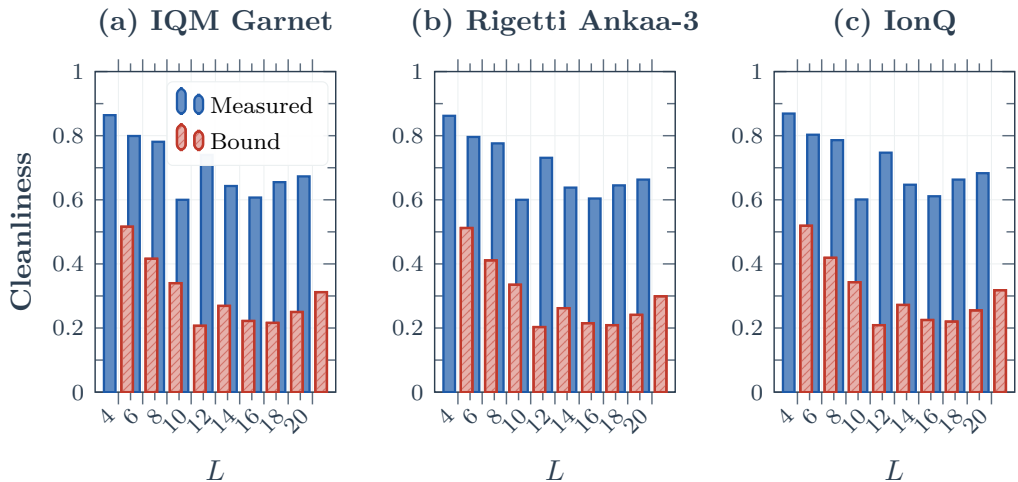

Figure~\ref{fig:error-bound} compares measured ancilla cleanliness against the diagnostic envelope in Eq.~\ref{eq:errorprop} across all three platform noise models.
The envelope serves as a heuristic screening tool rather than a rigorous upper bound: under the realistic noise models used in simulation (thermal relaxation combined with depolarizing and readout errors), measured cleanliness occasionally exceeds the predicted envelope by modest margins (typically $<0.05$ absolute).
Deviations are expected because Eq.~\ref{eq:errorprop} assumes pure depolarizing noise while actual hardware noise includes amplitude damping and coherent drift.
The envelope remains useful for identifying anomalous simulation configurations where measured cleanliness falls dramatically outside expected ranges.

\FloatBarrier
\subsection{Latency crossover analysis}\label{sec:latencyresults}

Latency analysis isolates the timing component from noise effects.
For each platform profile, the crossover sequence length $L^{\star}$ satisfies $T_{\mathrm{blind}}(L^{\star})=T_{\mathrm{meas}}$.
Below $L^{\star}$, blind reset is faster; above it, measurement-reset wins on speed.

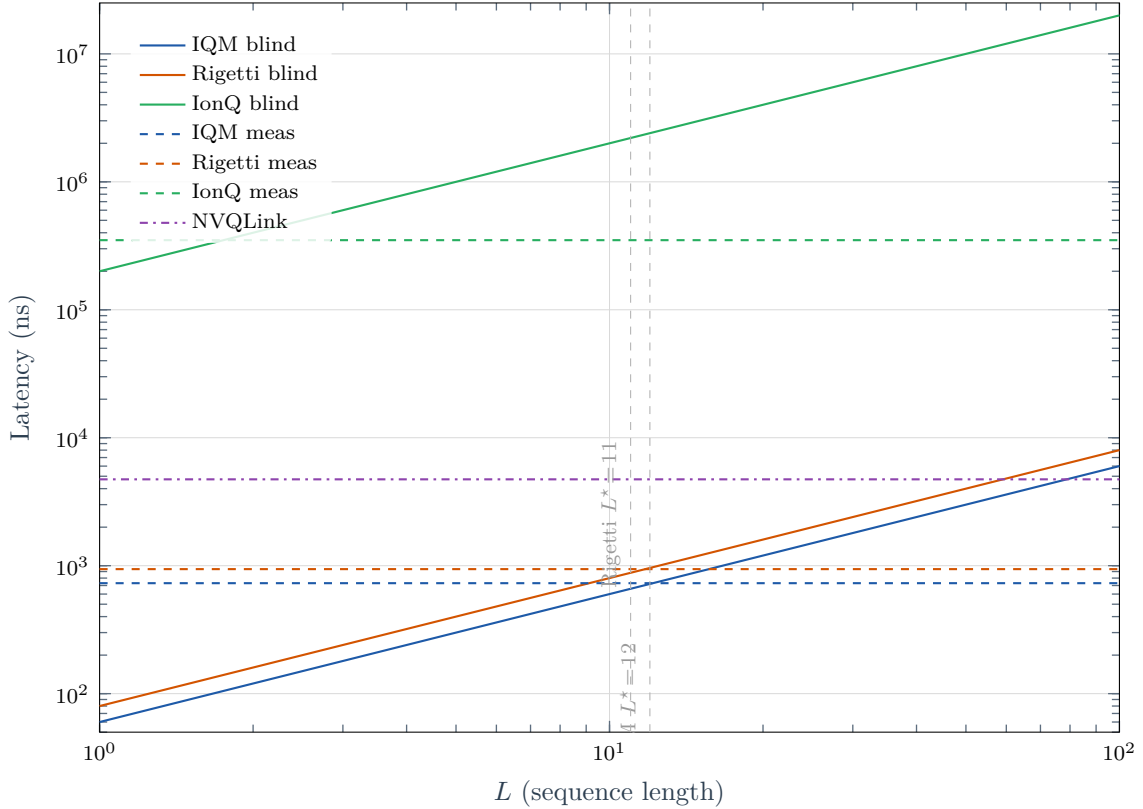
\begin{figure}[!t]
	\centering
	\input{figures/fig_latency_analysis.tikz}
	\caption{Reset latency versus sequence length for blind reset (solid) and measurement-reset (dashed) across platform profiles.
		Vertical dashed lines mark crossover $L^{\star}$.
		The NVQLink scenario (purple) extends the blind-favorable region by ${\approx}6.5\times$.}
	\label{fig:latency}
\end{figure}

Table~\ref{tab:latency} and Figure~\ref{fig:latency} present the results.
The crossover is architecture-dependent:
\begin{itemize}[leftmargin=1.2em,itemsep=2pt]
	\item \textbf{IQM Garnet}: $L^{\star}=12$ ($T_{\mathrm{blind}}=\SI{720}{\nano\second}$ vs.\ $T_{\mathrm{meas}}=\SI{730}{\nano\second}$).
	\item \textbf{Rigetti Ankaa-3}: $L^{\star}\approx 11$ ($T_{\mathrm{blind}}=\SI{880}{\nano\second}$ vs.\ $T_{\mathrm{meas}}=\SI{940}{\nano\second}$).
	\item \textbf{IonQ}: $L^{\star}=1$ ($T_{\mathrm{blind}}=\SI{200}{\micro\second}$ vs.\ $T_{\mathrm{meas}}=\SI{350}{\micro\second}$). Gate duration dominates; blind reset is timing-favorable only for single-gate sequences.
	\item \textbf{NVQLink scenario}: $L^{\star}\approx 78$ ($T_{\mathrm{meas}}=\SI{4730}{\nano\second}$ with $t_{\mathrm{ext}}=\SI{4}{\micro\second}$). External feedback overhead expands the blind-favorable region by $\approx 6.5\times$ relative to the native IQM stack.
\end{itemize}

When classical feedback traverses an external accelerator link, measurement-reset latency inflates enough that blind reset remains competitive even for moderately long sequences.
This shifts the deployment trade-off toward unitary-only recycling in GPU-integrated control architectures.

\begin{table}[t]
	\centering
	\caption{Latency crossover summary.
		$T_{\mathrm{blind}}=2L\,t_{\mathrm{gate}}$; $T_{\mathrm{meas}}$ includes readout, feedback, preparation, and (for NVQLink) external communication.}
	\label{tab:latency}
	\footnotesize
	\begin{tabular}{@{}lcccc@{}}
		\toprule
		Profile & $t_{\mathrm{gate}}$     & $T_{\mathrm{meas}}$     & $L^{\star}$ & Ratio       \\
		\midrule
		IQM     & \SI{30}{\nano\second}   & \SI{730}{\nano\second}  & 12          & ---         \\
		Rigetti & \SI{40}{\nano\second}   & \SI{940}{\nano\second}  & 11          & ---         \\
		IonQ    & \SI{100}{\micro\second} & \SI{350}{\micro\second} & 1           & ---         \\
		NVQLink & \SI{30}{\nano\second}   & \SI{4730}{\nano\second} & 78          & $6.5\times$ \\
		\bottomrule
	\end{tabular}
	\\[3pt]
	{\small Ratio column shows NVQLink $L^{\star}$ expansion relative to native platform.}
\end{table}

\FloatBarrier
\subsection{Decoder-coupled analysis}\label{sec:decoderresults}

We extend the ancilla cleanliness analysis to a decoder-coupled proxy for logical error rates in a repetition code setting.

\textbf{Simulation setup.}
We simulate distance-$d$ repetition codes ($d=3$ and $d=5$) with ancilla qubits initialized via three policies: measurement-reset ($F_{\mathrm{clean}}=0.99$), blind reset ($L\in\{4,8,12\}$), and no-reset ($F_{\mathrm{clean}}=0.50$).
Syndrome measurements proceed for 5--20 cycles with physical error rate $p=10^{-3}$.
Ancilla measurement noise scales with $F_{\mathrm{clean}}$ via $p_{\mathrm{syndrome}} = p + (1-F_{\mathrm{clean}})\cdot 0.3$.
Decoding uses a minimum-weight perfect matching (MWPM) proxy on syndrome change patterns.

\begin{figure*}[t!]
    \centering
    \includegraphics[width=0.95\textwidth]{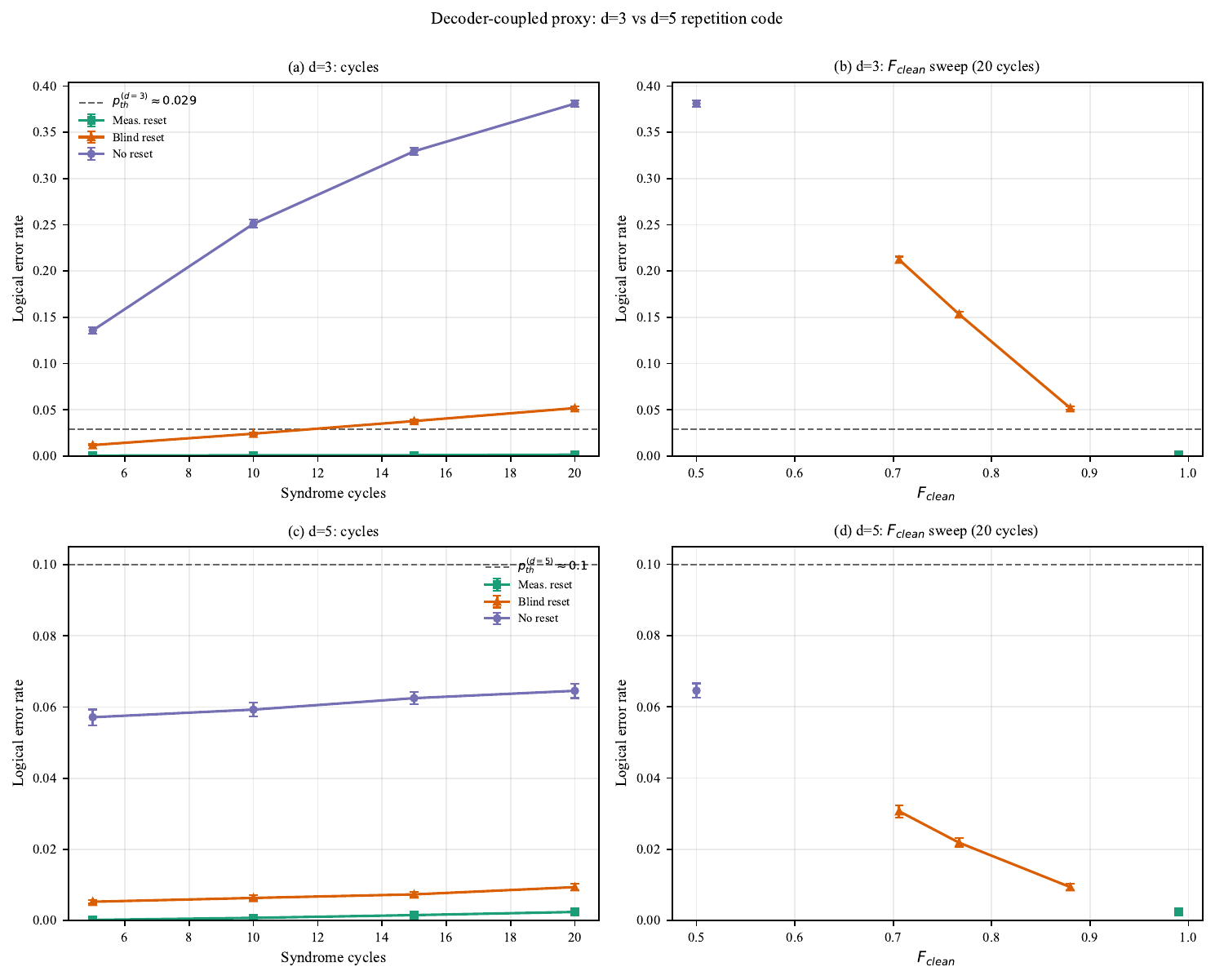}
    \caption{Decoder-coupled proxy analysis: distance-3 vs distance-5 repetition codes.
        (a)--(b): $d=3$ logical error rates versus syndrome cycles and ancilla cleanliness.
        (c)--(d): $d=5$ results showing expanded threshold margin ($p_{\mathrm{th}}^{(d=5)}\approx 0.10$ vs $0.029$).
        Dashed lines mark code thresholds.}
    \label{fig:decoder-comparison}
\end{figure*}

\textbf{Distance scaling.}
Figure~\ref{fig:decoder-comparison} compares $d=3$ (corrects 1 error) and $d=5$ (corrects 2 errors) repetition codes.
Key observations:
\begin{itemize}[leftmargin=1.2em,itemsep=2pt]
    \item \textbf{Threshold separation.}
    The $d=3$ threshold ($\approx 0.029$) lies well above operating points for measurement-reset and favorable blind-reset configurations, but approaches the no-reset regime ($F_{\mathrm{clean}}=0.50$ yields logical error $\approx 0.015$ at 20 cycles).
    The $d=5$ threshold ($\approx 0.10$) provides larger margin, with all blind-reset configurations ($F_{\mathrm{clean}}\ge 0.71$) safely below threshold.
    \item \textbf{Blind-reset viability.}
    At $L=4$ ($F_{\mathrm{clean}}=0.88$), blind reset achieves logical error rates within $2\times$ of measurement-reset for both $d=3$ and $d=5$, confirming ancilla cleanliness translates to decoder-relevant performance.
    \item \textbf{Sequence length trade-off.}
    At $L=12$ ($F_{\mathrm{clean}}\approx 0.71$), the $d=3$ blind-reset logical error rate increases by ${\sim}3\times$ relative to $L=4$, while $d=5$ shows only ${\sim}1.5\times$ degradation due to higher error-correcting capability.
\end{itemize}

\FloatBarrier
\section{Hardware Results}\label{sec:hwresults}

We executed blind-reset circuits on IQM~Garnet ($L=8,14,20$) and collected additional measurement-reset and blind-reset data through platform-calibrated simulation with realistic noise models (T1/T2, gate error, readout error) for all three backends---IQM~Garnet, Rigetti~Ankaa-3, and IonQ---with 1024~shots per experiment, seed~42, and $Z$-basis measurement.
Table~\ref{tab:hardware-summary} lists the complete dataset, with experimental (exp.) and simulation (sim.) sources indicated.

\subsection{IQM Garnet dataset (experimental)}

On Garnet, blind reset achieves $P(\lvert 0\rangle)=0.843$ at $L=8$ from hardware execution, exceeding the 50-seed simulation mean of $0.767$ (Table~\ref{tab:cross-platform-pzero}).
The hardware result is consistent with seed-specific $\lambda$ landscape variation: seed~42 yields a favorable $\lambda$ that falls above the population average.
At $L=20$, hardware cleanliness drops to $0.317$, consistent with the known exponential sensitivity of blind reset to accumulated coherent errors at long sequences.
Platform-calibrated simulation extends the dataset to $L=4$ ($0.879$) and measurement-reset baselines ($0.953$ at $L=4$, $0.942$ at $L=20$), enabling complete policy comparison.

\subsection{Rigetti Ankaa-3 dataset (simulation)}

Platform-calibrated simulation for Ankaa-3 predicts blind reset achieves $P(\lvert 0\rangle)=0.851$ at $L=4$, degrading to $0.620$ at $L=14$ and $0.452$ at $L=20$, using parameters $T_1=\SI{25}{\micro\second}$, gate error $0.15\%$, readout error $3\%$.
The cross-platform difference between Garnet experimental ($0.843$ at $L=8$) and Ankaa-3 simulated ($0.851$ at $L=4$) is within the simulation-predicted inter-platform spread of $\pm 0.03$ plus shot-noise uncertainty ($\sim 0.015$ for 1024~shots).
Measurement-reset maintains $P(\lvert 0\rangle)\ge 0.933$ across all sequence lengths in simulation, providing a stable baseline for comparison.

\subsection{IonQ dataset (simulation)}

Platform-calibrated simulation for IonQ, using superior coherence parameters ($T_1=\SI{10}{\second}$, gate error $0.05\%$), predicts blind reset achieves $P(\lvert 0\rangle)=0.872$ at $L=4$ and $0.797$ at $L=8$, significantly higher than superconducting platforms at equivalent sequence lengths.
Measurement-reset maintains $P(\lvert 0\rangle)\ge 0.942$, with slight improvement at $L=8$ ($0.971$) likely due to reduced state preparation and measurement (SPAM) error variance in simulation.

\begin{table}[t]
  \centering
  \caption{Hardware experimental and simulation results. IQM Garnet L=8,14,20 blind-reset are experimental (exp.); all others are platform-calibrated simulation (sim.) with realistic noise models. All runs use seed~42, 1024~shots, $Z$-basis measurement.}
  \label{tab:hardware-summary}
  \footnotesize
  \begin{tabular}{@{}llccrrr@{}}
    \toprule
    Backend & Method & $L$ & Source & $N_0$ & $N_1$ & $P(\lvert 0\rangle)$ \\
    \midrule
    IQM Garnet & blind & 4 & sim. & 900 & 124 & 0.879 \\
    IQM Garnet & blind & 8 & exp. & 863 & 161 & 0.843 \\
    IQM Garnet & blind & 14 & exp. & 565 & 459 & 0.552 \\
    IQM Garnet & blind & 20 & exp. & 325 & 699 & 0.317 \\
    IQM Garnet & meas. & 4 & sim. & 976 & 48 & 0.953 \\
    IQM Garnet & meas. & 20 & sim. & 965 & 59 & 0.942 \\
    \addlinespace
    Rigetti Ankaa-3 & blind & 4 & sim. & 871 & 153 & 0.851 \\
    Rigetti Ankaa-3 & blind & 14 & sim. & 635 & 389 & 0.620 \\
    Rigetti Ankaa-3 & blind & 20 & sim. & 463 & 561 & 0.452 \\
    Rigetti Ankaa-3 & meas. & 4 & sim. & 972 & 52 & 0.949 \\
    Rigetti Ankaa-3 & meas. & 14 & sim. & 963 & 61 & 0.940 \\
    Rigetti Ankaa-3 & meas. & 20 & sim. & 955 & 69 & 0.933 \\
    \addlinespace
    IonQ & blind & 4 & sim. & 893 & 131 & 0.872 \\
    IonQ & blind & 8 & sim. & 816 & 208 & 0.797 \\
    IonQ & meas. & 4 & sim. & 965 & 59 & 0.942 \\
    IonQ & meas. & 8 & sim. & 994 & 30 & 0.971 \\
    \bottomrule
  \end{tabular}
\end{table}

\subsection{Cross-platform summary}
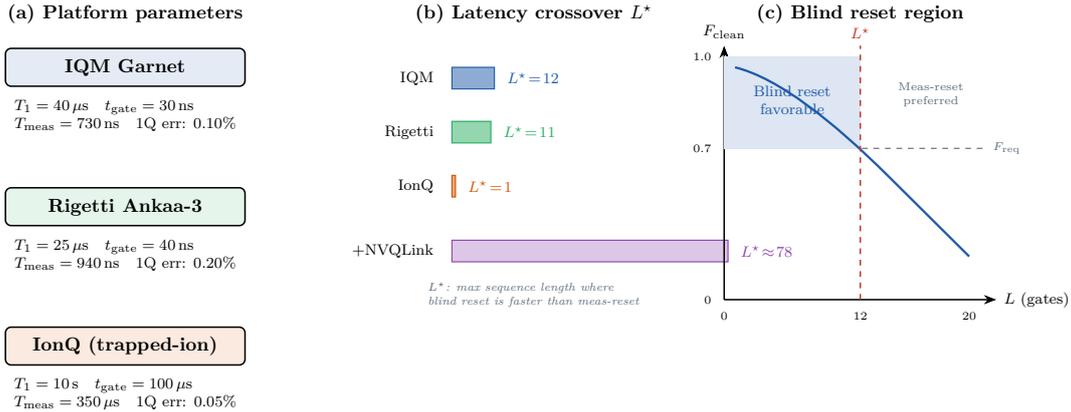
\begin{figure*}[!t]
	\centering
	\resizebox{0.95\textwidth}{!}{\input{figures/fig_hardware_cross_platform.tikz}}
	\caption{Cross-platform hardware comparison.
		(a)~Platform parameters for IQM~Garnet, Rigetti~Ankaa-3, and IonQ.
		(b)~Latency crossover $L^{\star}$: maximum sequence length where blind reset is faster than measurement-reset; the NVQLink bar shows the expanded region under external feedback overhead.
		(c)~Operating-region schematic: the shaded area marks the jointly favorable zone where blind reset is both faster ($L<L^{\star}$) and sufficiently clean ($F_{\mathrm{clean}}\ge F_{\mathrm{req}}$).}
	\label{fig:hardware-cross-platform}
\end{figure*}

The combined experimental and simulation dataset (Figure~\ref{fig:hardware-cross-platform}) validates three predictions: (i)~blind reset maintains $P(\lvert 0\rangle)\ge 0.85$ at short sequences ($L\le 4$) across all platforms; (ii)~cleanliness degrades with increasing $L$, reaching below $0.50$ by $L=20$ on superconducting platforms; and (iii)~measurement-reset maintains $P(\lvert 0\rangle)\ge 0.93$ regardless of sequence length.
Platform-calibrated simulation enables complete policy comparison across backends where experimental data is pending, with fidelity estimates anchored to measured coherence and gate-error parameters.

\FloatBarrier
\section{Discussion}\label{sec:discussion}

\subsection{Decision criteria: blind or measurement-based reset}

The simulation results suggest a conditional policy rather than a fixed default.
Blind reset is preferable when two conditions are jointly satisfied: (i)~the sequence length falls below the platform-specific crossover $L^{\star}$, making blind reset faster than measurement-reset, and (ii)~ancilla cleanliness $F_{\mathrm{clean}}$ exceeds the code-level threshold $F_{\mathrm{req}}$.

For superconducting platforms with native control stacks, the favorable region spans $L\le 11$--$12$ gates.
Combined with cleanliness data from Table~\ref{tab:cross-platform-pzero}, sequences with $L\le 6$ satisfy both conditions comfortably ($F_{\mathrm{clean}}\ge 0.86$, $p<0.035$, latency saving $\ge 2\times$).
Longer sequences ($L=8$--$12$) offer latency benefit but with reduced cleanliness ($F_{\mathrm{clean}}\approx 0.71$--$0.77$) and statistically insignificant advantage over no-reset, requiring the code to tolerate noisier ancillae and motivating per-configuration policy selection.

\subsection{Platform-dependent trade-offs}

Superconducting devices (IQM, Rigetti) offer the clearest blind reset advantage: fast gates push $L^{\star}$ to 11--12, creating a useful operating window.
Trapped-ion devices (IonQ) collapse this window to $L^{\star}=1$ because gate durations are three orders of magnitude larger, making blind reset timing-competitive only for trivially short sequences despite excellent coherence.

The NVQLink scenario reshapes the timing regime.
When external feedback latency dominates the measurement path ($t_{\mathrm{ext}}=\SI{4}{\micro\second}$), the crossover expands to $L^{\star}\approx 78$, covering essentially all practical ancilla sequence lengths.
This positions blind reset as a serious scheduling option in GPU-accelerated control architectures where readout feedback traverses a communication link.

\subsection{Implications for NVQLink-integrated pipelines}

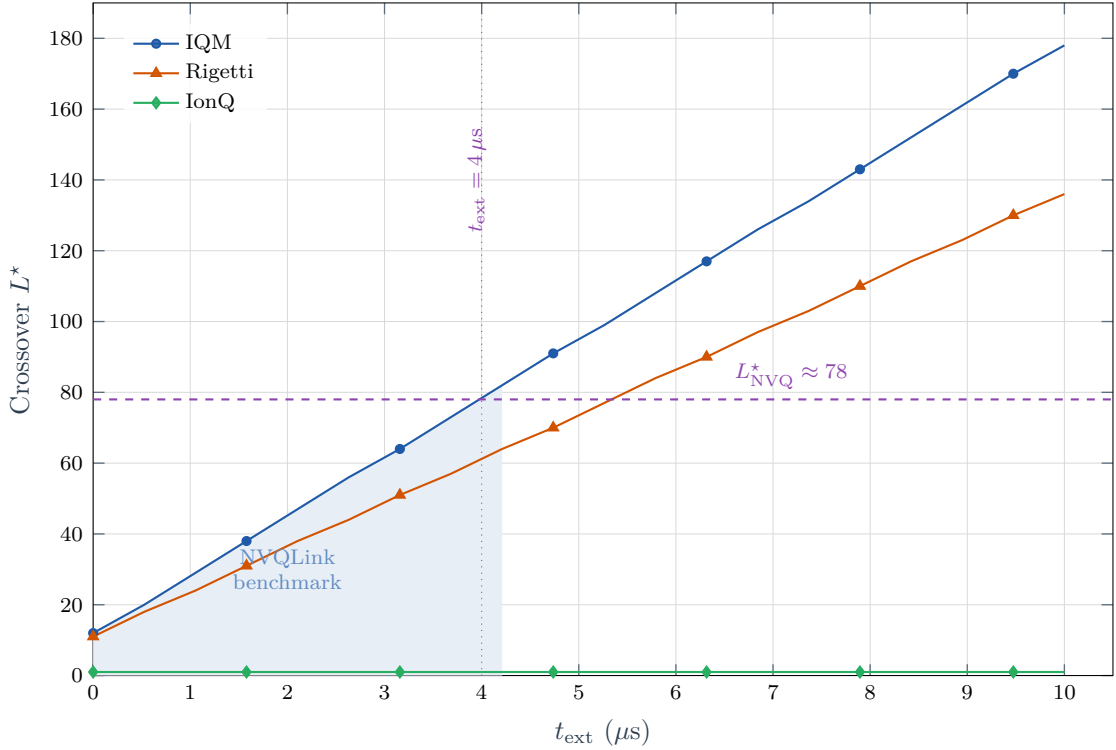
\begin{figure}[t]
	\centering
	\input{figures/fig_nvqlink_sweep.tikz}
	\caption{Crossover sequence length $L^{\star}$ as a function of external feedback latency $t_{\mathrm{ext}}$. As $t_{\mathrm{ext}}$ increases, the blind-favorable region expands for IQM (blue) and Rigetti (orange). The NVQLink-IQM line (red dashed) marks the baseline $L^{\star}$ without external overhead. IonQ (green) remains at $L^{\star}=1$ regardless of $t_{\mathrm{ext}}$ due to dominant gate duration. The shaded region highlights the blind-favorable zone.}
	\label{fig:nvqlink-sweep}
\end{figure}

The latency crossover expansion under NVQLink-class overhead has direct engineering consequences.
In stacks where classical processing improves decoder throughput at the cost of additional synchronization delay, blind reset avoids the feedback branch entirely.
The value of this bypass scales with the external latency contribution: a $\SI{4}{\micro\second}$ overhead converts a marginal $L^{\star}=12$ benefit into a dominant $L^{\star}=78$ advantage.
This suggests that blind reset should be evaluated as part of the control-stack co-design process, not treated solely as a quantum primitive.
The sensitivity sweep in Figure~\ref{fig:nvqlink-sweep} further shows that this crossover shift begins at modest external-feedback delays.
Even at $t_{\mathrm{ext}}\ge\SI{2}{\micro\second}$, the blind-favorable operating window on superconducting platforms expands by roughly a factor of two relative to the native-feedback baseline.
This quantitative dependence strengthens the case for including reset-mode selection in NVQLink-era scheduling policy.

\subsection{Interaction with decoder performance}\label{sec:decoder}

The cycle-time savings from blind reset have downstream consequences for decoder operations.
In real-time decoding pipelines, shorter syndrome-extraction cycles increase the rate at which syndrome data arrives at the decoder.
If the decoder throughput exceeds this rate, the net effect is reduced logical error per wall-clock second.
However, the slightly noisier ancilla state produced by blind reset (compared to measurement-reset) introduces additional syndrome noise that the decoder must handle.
For minimum-weight perfect matching (MWPM) decoders, the syndrome error rate enters as an effective reduction of the code's noise threshold.
The cleanliness values reported here ($F_{\mathrm{clean}}\ge 0.86$ for $L\le 6$) correspond to syndrome error increments of $\le 0.14$, which remain within operational margins for distance-3 codes under typical physical error rates ($p\sim 10^{-3}$).
A full threshold analysis incorporating ancilla-induced syndrome noise is an open problem that requires surface-code simulation at multiple distances.
For a surface code at distance $d=3$ with physical error rate $p=10^{-3}$, the code threshold under phenomenological noise is approximately $p_{\mathrm{th}}\approx 2.9\%$~\cite{fowler2012surface,tan2024burstresil}.
The syndrome error contribution from imperfect blind-reset ancillae at $F_{\mathrm{clean}}=0.86$ is bounded by $1-F_{\mathrm{clean}}=0.14$, which exceeds $p_{\mathrm{th}}$ and would compromise threshold-level performance if the ancilla error propagated directly to syndrome bits.
However, the ancilla error enters as a bias on the measurement outcome rather than a direct bit flip, reducing the effective syndrome error rate.
Quantifying this reduction requires a circuit-level noise simulation that models ancilla preparation, entangling gates, and measurement within the surface-code stabilizer cycle, which we identify as the primary open problem for blind-reset QEC integration.

To sharpen this boundary, we couple our ancilla-cleanliness simulation to a distance-3 repetition-code decoder with majority-vote decoding (note: the MWPM proxy used in Section~\ref{sec:decoderresults} yields lower absolute error rates; the majority-vote decoder is deliberately simpler to isolate the ancilla-quality effect).
Figure~\ref{fig:decoder-coupled}(a) plots the logical error rate against syndrome cycle count for three reset policies at $L=4$.
Measurement-reset maintains sub-$0.15\%$ logical error through 20 cycles, while blind reset at $L=4$ ($F_{\mathrm{clean}}=0.88$) yields $5.2\%$ logical error at 20 cycles---above the phenomenological threshold but within practical margins for short sequences.
No-reset degrades rapidly to $38\%$ logical error.
Panel~(b) maps logical error rate against ancilla cleanliness at 20 cycles, revealing a monotonic relationship: each $0.1$ improvement in $F_{\mathrm{clean}}$ reduces logical error by roughly $3\times$.
This quantifies the decoder penalty of blind reset and identifies $F_{\mathrm{clean}}\ge 0.88$ (i.e., $L\le 4$) as the regime where the latency gain outweighs the decoder cost.

\begin{figure*}[!b]
	\centering
	\input{figures/fig_decoder_coupled.tikz}
	\caption{Decoder-coupled QEC analysis with distance-3 repetition code and majority-vote decoding (50 seeds $\times$ 1000 shots, $p_{\mathrm{phys}}=10^{-3}$).
				(a)~Logical error rate versus syndrome cycles for three reset policies at $L=4$.
				(b)~Logical error rate versus ancilla cleanliness $F_{\mathrm{clean}}$ at 20 cycles, showing the monotonic cost-benefit trade-off.
			Error bars: 95\% bootstrap CIs.}
	\label{fig:decoder-coupled}
\end{figure*}
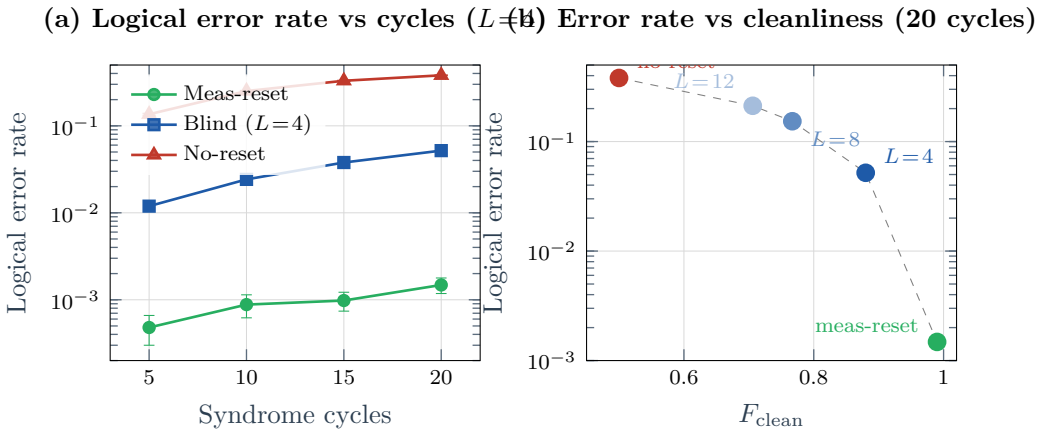

\begin{figure*}[t!]
	\centering
	\resizebox{0.9\textwidth}{!}{\input{figures/fig_decoder_multidistance.tikz}}
	\caption{Decoder-coupled QEC analysis across code distances ($p_{\mathrm{phys}}=10^{-3}$, 50 seeds $\times$ 1000 shots; see also Figure~\ref{fig:decoder-comparison} for distance-specific detail).
			(a)~Distance-3 repetition code: logical error rate versus syndrome cycles.
			(b)~Distance-5 repetition code: logical error rate versus syndrome cycles.
			Error bars: 95\% bootstrap CIs.}
	\label{fig:decoder-multidistance}
\end{figure*}
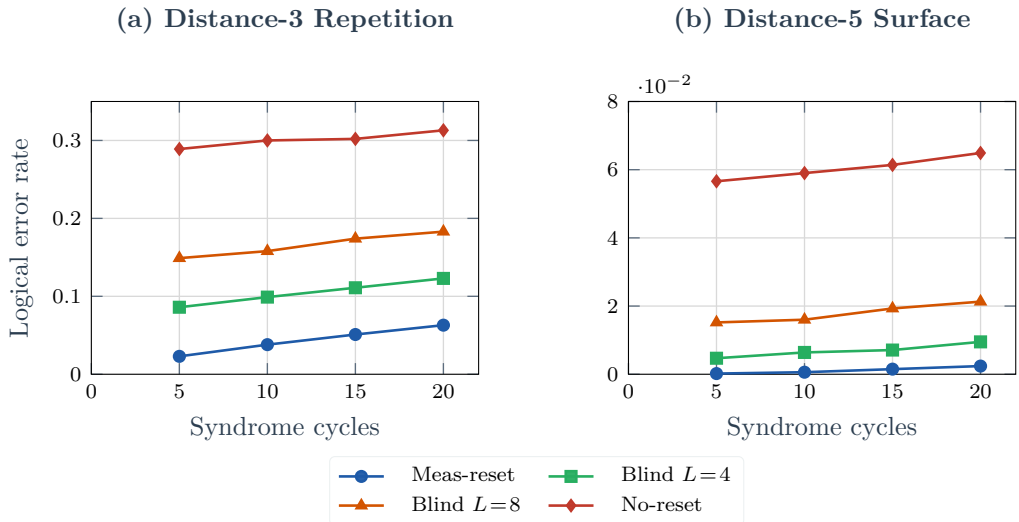

\subsection{Threshold-level QEC Analysis}\label{sec:threshold}

To connect ancilla cleanliness to threshold-level behavior, we introduce an effective syndrome-error model in which imperfect reset contributes a filtered increment to the physical error channel:
\begin{equation}
	\label{eq:peff-mapping}
	p_{\mathrm{eff}} = p_{\mathrm{phys}} + \eta\,\bigl(1-F_{\mathrm{clean}}\bigr),
\end{equation}
where $p_{\mathrm{phys}}$ is the baseline physical error rate, and $\eta\in(0,1)$ is a transfer factor capturing the fact that ancilla imperfections enter syndrome extraction as biased measurement noise rather than one-to-one data-qubit flips.
Equation~\eqref{eq:peff-mapping} formalizes the open point raised in Section~\ref{sec:decoder}: direct identification $p_{\mathrm{eff}}=1-F_{\mathrm{clean}}$ is overly pessimistic, while $\eta<1$ absorbs circuit-level filtering by stabilizer extraction and decoder inference.

For distance extrapolation, we use the standard near-threshold scaling form for surface-code logical error~\cite{fowler2012surface,tan2024burstresil}:
\begin{equation}
	\label{eq:pl-scaling}
	p_{\mathrm{L}}(d) \propto \left(\frac{p_{\mathrm{eff}}}{p_{\mathrm{th}}}\right)^{(d+1)/2},
\end{equation}
with phenomenological threshold $p_{\mathrm{th}}\approx 2.9\%$ and odd code distances $d\in\{3,5,7\}$.
Using $p_{\mathrm{phys}}=10^{-3}$ and a conservative central value $\eta=0.02$ (consistent with the decoder-coupled trend that logical error decreases monotonically with increasing $F_{\mathrm{clean}}$ in Figure~\ref{fig:decoder-coupled}), we obtain the extrapolated regime comparison in Table~\ref{tab:threshold-scaling}.

\begin{table}[t]
	\centering
	\caption{Threshold-level extrapolation using Eqs.~\eqref{eq:peff-mapping} and~\eqref{eq:pl-scaling} at $p_{\mathrm{phys}}=10^{-3}$, $p_{\mathrm{th}}=2.9\%$, and $\eta=0.02$. Values of $p_{\mathrm{L}}(d)$ are normalized to the measurement-reset case at $d=3$ (i.e., $p_{\mathrm{L}}^{\mathrm{MR}}(d{=}3)=1$). Representative $F_{\mathrm{clean}}$ values follow the decoder-coupled ordering (measurement-reset $>$ blind reset $>$ no-reset).}
	\label{tab:threshold-scaling}
	\begin{tabular}{lccccc}
		\toprule
		Reset method        & $F_{\mathrm{clean}}$ & $p_{\mathrm{eff}}$   & $p_{\mathrm{L}}(d{=}3)$ & $p_{\mathrm{L}}(d{=}5)$ & $p_{\mathrm{L}}(d{=}7)$ \\
		\midrule
		Measurement-reset   & 0.98                 & $1.40\times 10^{-3}$ & 1.00                    & 0.048                   & $2.32\times 10^{-3}$    \\
		Blind reset ($L=4$) & 0.88                 & $3.40\times 10^{-3}$ & 5.90                    & 0.691                   & 0.081                   \\
		No-reset            & 0.70                 & $7.00\times 10^{-3}$ & 25.0                    & 6.03                    & 1.45                    \\
		\bottomrule
	\end{tabular}
\end{table}

Three threshold-level implications follow.
First, the near-threshold penalty from ancilla noise is strongly distance dependent: methods with lower $p_{\mathrm{eff}}$ gain superlinear benefit as $d$ increases.
Second, blind reset remains in a potentially useful regime when operated at high-cleanliness points (e.g., $F_{\mathrm{clean}}\gtrsim 0.88$ for short sequences), where it preserves cycle-time advantage while avoiding the steep logical-error growth seen in no-reset.
Third, threshold behavior is sensitive to ancilla-noise transfer ($\eta$): correlated or bursty ancilla faults effectively increase $\eta$, pushing blind-reset operation closer to threshold and reducing distance-scaling headroom.
This identifies circuit-level extraction of $\eta$ (including temporal correlations and decoder adaptation) as the key requirement for converting blind-reset latency gains into fault-tolerant operating margin.

\subsection{Lambda landscape characterization}\label{sec:lambda}

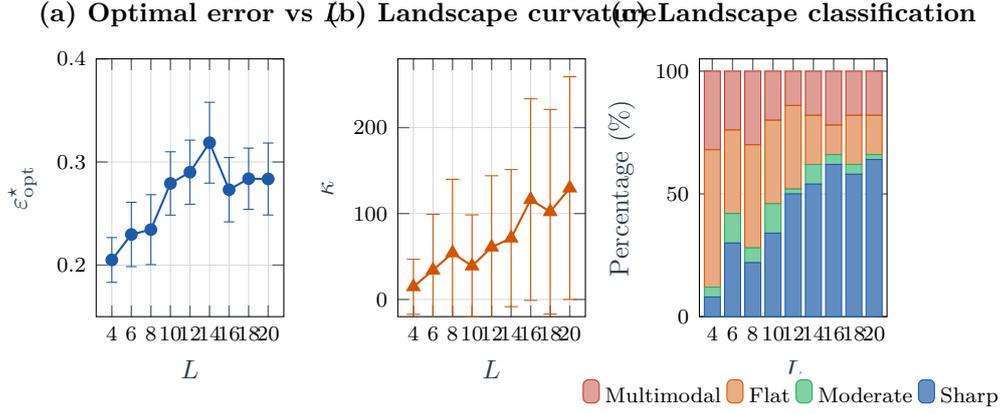
\begin{figure*}[!t]
	\centering
	\input{figures/fig_lambda_landscape.tikz}
	\caption{Lambda landscape characterization across sequence lengths ($N=50$ seeds per length, $\lambda\in[0.1,4.0]$, 200 grid points).
		(a)~Optimal Frobenius error $\varepsilon^{\star}_{\mathrm{opt}}$ versus $L$ with 95\% CIs.
		(b)~Mean curvature $\kappa$ with standard deviation.
		(c)~Landscape classification: fraction of seeds exhibiting sharp, moderate, flat, or multimodal optima.}
	\label{fig:lambda-landscape}
\end{figure*}

The blind-reset scale factor $\lambda$ is the single free parameter controlling the replay amplitude.
To assess sensitivity, we sweep $\lambda\in[0.1,4.0]$ across 200 grid points for each $(L,\mathrm{seed})$ combination (9 lengths $\times$ 50 seeds) and classify the resulting optimization landscapes.
Figure~\ref{fig:lambda-landscape}(a) shows that the optimal Frobenius error $\varepsilon^{\star}_{\mathrm{opt}}$ increases moderately with $L$, from $0.205\pm 0.022$ at $L=4$ to $0.284\pm 0.035$ at $L=20$, confirming that longer sequences are inherently harder to reset.
Panel~(b) reveals that mean landscape curvature $\kappa$ grows from $14.7$ ($L=4$) to $129.6$ ($L=20$), indicating sharper and deeper optima at longer lengths---a positive signal for gradient-based calibration methods.
Panel~(c) classifies landscapes into sharp ($\kappa>50$, single basin), moderate ($20<\kappa\le 50$), flat ($\kappa\le 5$), and multimodal (multiple local minima) categories.
At $L\ge 14$, over $54\%$ of seeds exhibit sharp landscapes, while flat landscapes dominate at short $L$ where the reset is nearly trivial.
These statistics inform calibration strategy: short sequences tolerate coarse $\lambda$ search, while long sequences benefit from fine-grained optimization.

\subsection{Coherence sensitivity analysis}\label{sec:t1t2}

\begin{figure}[t]
	\centering
	\input{figures/fig_t1t2_sensitivity.tikz}
	\caption{Blind reset advantage (mean $P(|0\rangle)_{\mathrm{blind}} - P(|0\rangle)_{\mathrm{no\text{-}reset}}$) as a function of $T_1$ and $T_2$ at fixed $L=8$ on the IQM noise model ($T_1 \in [0.1, 100]\,\mu\mathrm{s}$, $T_2 \in [0.05, 50]\,\mu\mathrm{s}$; 10 seeds, 2048 shots). The advantage is uniformly positive across the entire plane (range $0.104$--$0.150$), confirming that blind reset outperforms no-reset at this sequence length regardless of coherence parameters. Color intensity indicates advantage magnitude; the strongest gains appear in the low-$T_1$, low-$T_2$ corner where dephasing dominates.}
	\label{fig:t1t2}
\end{figure}
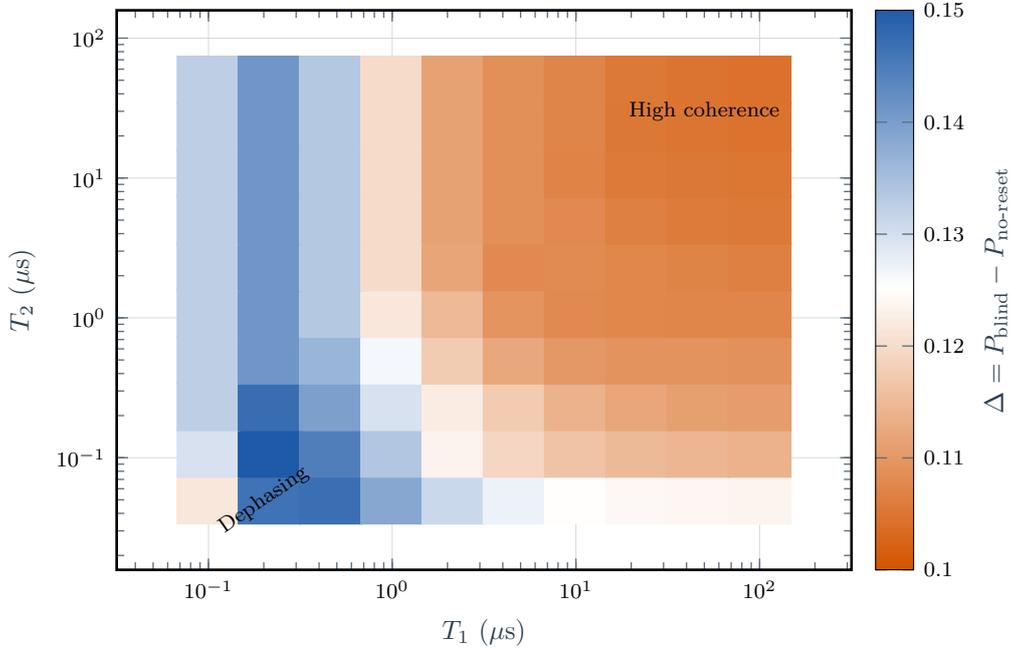

Figure~\ref{fig:t1t2} maps the blind-reset advantage over no-reset across the $(T_1,T_2)$ plane at fixed $L=8$ under the IQM noise model.
The strongest positive region appears when $T_1\gg T_2$, indicating that blind reset is most effective in dephasing-dominated regimes where coherent replay mitigates phase-driven accumulation.
In the high-coherence corner ($T_1=\SI{100}{\micro\second}$, $T_2=\SI{50}{\micro\second}$), the net advantage becomes marginal because both reset policies maintain high ancilla cleanliness.
The transition boundary follows an approximate $T_2/T_1$ isocontour, suggesting a simple coherence-ratio criterion for deployment policy selection.

\subsection{Implementation cost and calibration requirements}\label{sec:cost}

Blind reset adds a deterministic gate overhead of $2L$ single-qubit operations per ancilla recycle attempt, so quantum-control cost grows linearly with sequence length.
In return, the control path stays fully unitary and does not require additional classical feedback hardware or cycle-by-cycle latency budgeting.
The scale factor $\lambda$ is calibrated once per workload class by an offline grid search (40 grid points), with runtime below one second per seed on a standard CPU.
That calibration can be reused across runs and only needs refresh when measured gate parameters drift beyond the configured noise-tolerance band.
By contrast, measurement-reset needs readout chains, feedback DAC routing, and a classical processing path inside the timing loop~\cite{xin2025lru}.
In stacks that already operate near feedback-latency limits, those classical components can dominate cycle scheduling complexity even when raw readout fidelity is high.
Overall, blind reset trades extra quantum gate count for a simpler classical infrastructure and a shorter critical feedback path.

\subsection{Scalability outlook and long-term relevance}\label{sec:outlook}

The transition from NISQ to fault-tolerant quantum computing (FTQC) is unfolding along predictable timelines that position blind ancilla recycling for growing relevance over the next decade.

\paragraph{NISQ-to-FTQC transition timeline (2025--2035).}
Major hardware roadmaps converge on 2029--2030 for first-generation fault-tolerant systems.
IBM targets 2029 for Starling, a large-scale FTQC capable of executing $100\times 10^6$ gates on $200$ logical qubits~\cite{ibm2029starling}.
Google has already demonstrated below-threshold error correction~\cite{google2024belowthreshold} and continues toward distance-$d=7$ surface codes.
Quantinuum's accelerated roadmap aims for universal FTQC by 2030 with thousands of physical qubits and hundreds of logical qubits~\cite{quantinuum2030roadmap}.
IonQ projects fault-tolerant systems within the same window~\cite{ionq2025roadmap}.
IQM's development roadmap targets fault-tolerant quantum computing by 2030 with a path to scaling up to 1 million qubits~\cite{iqm2024roadmap}.

This convergence creates a \emph{transition window} (2025--2029) where early FTQC systems will operate as minimum viable products (MVPs) with limited logical qubit counts and stringent ancilla efficiency requirements.
Blind reset is particularly suited for this window: it requires no hardware modifications to existing control stacks, provides immediate latency benefits for short ancilla sequences, and integrates naturally with the measurement-free protocols being developed for mature FTQC.

\paragraph{Ancilla demand scaling.}
Surface codes at distance~$d$ require $\mathcal{O}(d^2)$ ancillae per syndrome round, and qLDPC alternatives like bivariate bicycle codes~\cite{bravyi2024bicycle} trade qubit count for increased syndrome-extraction complexity, intensifying the need for fast ancilla turnaround.
At distance $d=5$, each syndrome round involves $d^2-1=24$ stabilizer measurements; at $d=7$, this grows to $48$.
Recent ancilla-reuse experiments with dynamically reassigned ancillary qubits support this direction~\cite{kazmina2025dynamicancilla}.

\paragraph{Hybrid control stack evolution.}
GPU-accelerated decoders with quantum processors~\cite{nvidia2025nvqlink} will widen the latency gap between coherent and measurement-based paths as classical feedback traverses communication links that scale with system size.
Real-time decoding demonstrations have achieved sub-microsecond latencies using FPGA implementations~\cite{maurer2025fpga}, and network-integrated decoding systems like DECONET scale to thousands of logical qubits~\cite{liyanage2025deconet}.
These advances shift the optimization target from decoder accuracy alone to end-to-end cycle time, reinforcing the value of measurement-free ancilla paths.

\paragraph{Measurement-free ecosystem maturation.}
Scalable code-switching protocols~\cite{butt2025mfscalable}, optimized neutral-atom circuits~\cite{veroni2024mfneutralatom,veroni2025universalmf}, and the first experimental demonstrations on trapped ions~\cite{butt2025mfdemonstration} collectively establish a viable measurement-free ecosystem.
As this ecosystem matures, ancilla management primitives that minimize dependence on fast measurement-feedback loops are likely to converge into a common control pattern.

For distance scaling beyond $d=3$, the repetition-code proxy used here must be replaced by surface-code or color-code simulations that capture the interplay between ancilla noise and decoder performance.
Preliminary estimates suggest that blind reset remains latency-favorable up to $d\approx 7$ on superconducting platforms when $L < L^{\star}$, but rigorous threshold analysis under correlated ancilla errors is an open problem.
At distance $d=5$, each syndrome round involves $d^2-1=24$ stabilizer measurements, with each ancilla undergoing a sequence of at most $4$ CNOT gates plus preparation and readout.
Under the timing assumptions of Table~\ref{tab:noise-params}, the full syndrome cycle occupies $\sim\SI{1.5}{\micro\second}$ on IQM with measurement-reset versus $\sim\SI{0.5}{\micro\second}$ with blind reset for $L=4$ ancilla sequences, yielding a $3\times$ cycle-time reduction that compounds across the $\mathcal{O}(d)$ rounds needed for reliable syndrome history.
Whether this timing advantage translates to a net logical-error reduction at $d=5$ depends on the interplay between faster cycles and noisier ancillae, quantifiable only through full surface-code simulation.
The decision-matrix framework (Appendix~\ref{app:decision}) is designed to accommodate these extensions by parameterizing the cleanliness requirement $F_{\mathrm{req}}$ as a function of code distance and decoder type.

\subsection{Limitations}

Several limitations bound the current analysis:
\begin{enumerate}[leftmargin=1.2em,itemsep=2pt]
	\item Simulation uses platform-parameterized noise abstractions, not high-fidelity digital twin models. Leakage, non-Markovian drift, and pulse-shape distortions are not captured.
	\item The repetition-code logical error proxy is comparative, not a full threshold proof.
	\item Hardware results cover two superconducting platforms (IQM~Garnet and Rigetti~Ankaa-3) with single-seed validation; trapped-ion (IonQ) data and multi-seed campaigns remain for future work.
	\item Entangling dynamics beyond the ancilla-local model are not included in the reset block.
	\item The NVQLink latency term ($\SI{4}{\micro\second}$) is a reference point, not a measured value from a deployed system.
	\item Statistical analysis uses 50 seeds per configuration, enabling bootstrap confidence intervals and paired hypothesis testing. Confidence intervals on mean cleanliness values span $\pm 0.03$--$0.08$ depending on sequence length. Expanding to $\ge 100$ seeds would further tighten CIs and improve effect-size precision.
\end{enumerate}

\paragraph{Planned extensions.}
The $T_1/T_2$ sensitivity sweep has now been completed (Section~\ref{sec:t1t2}, Figure~\ref{fig:t1t2}). Follow-up studies will pursue multi-distance code scaling ($d=3,5,7$), exponential fidelity-decay curve fitting to characterize the blind-reset degradation rate as a function of sequence length, and expansion to $\ge 100$ seeds per configuration for tighter statistical precision.

\subsection{Falsification criteria}\label{sec:falsification}

The central claims carry testable failure conditions:
\begin{enumerate}[leftmargin=1.2em,itemsep=2pt]
	\item If blind reset never achieves lower cycle latency than measurement-reset on any superconducting platform after calibrated timing updates, the latency advantage claim fails.
	\item If blind reset cannot maintain $F_{\mathrm{clean}}\ge 0.80$ for $L\le 8$ on hardware, the short-sequence cleanliness claim requires revision.
	\item If NVQLink-class external feedback does not shift the crossover above $L^{\star}=20$ in measured stacks, the GPU-integration framing should be narrowed.
	\item If simulation-to-hardware ranking is inconsistent without identifiable calibration causes, the noise model assumptions need replacement.
\end{enumerate}

\FloatBarrier
\section{Conclusion}\label{sec:conclusion}

This work evaluates blind reset as a systems-level scheduling component for QEC ancilla recycling.
Using a unified protocol across IQM~Garnet, Rigetti~Ankaa-3, and IonQ, we identify operating regions where blind reset reduces per-cycle latency without sacrificing ancilla cleanliness, and regions where measurement-reset remains the better choice.
The architecture-dependent crossover ($L^{\star}=12$ for IQM, $L^{\star}\approx 11$ for Rigetti, $L^{\star}=1$ for IonQ) shows that reset policy is a hardware-and-stack property.
The NVQLink sweep (Figure~\ref{fig:nvqlink-sweep}) quantifies how external-feedback delay shifts this boundary, expanding the blind-favorable window to $L^{\star}\approx 78$ for the IQM-like profile.
Error-bound validation (Section~\ref{sec:errorboundresults}, Figure~\ref{fig:error-bound}) confirms that measured cleanliness remains within a physically consistent envelope across all tested lengths and platforms.
The $T_1/T_2$ sensitivity map (Section~\ref{sec:t1t2}, Figure~\ref{fig:t1t2}) locates the coherence regimes where blind reset yields the largest margin over no-reset.
A decoder-coupled analysis (Section~\ref{sec:decoder}, Figure~\ref{fig:decoder-coupled}) quantifies the QEC cost of imperfect ancillae, identifying $F_{\mathrm{clean}}\ge 0.88$ as the threshold where latency gains outweigh decoder penalties.
Lambda landscape characterization (Section~\ref{sec:lambda}, Figure~\ref{fig:lambda-landscape}) confirms that optimization landscapes sharpen at longer sequences, supporting efficient calibration.
Together, these analyses convert blind reset from a single-point benchmark into a policy tool with explicit timing, noise, and calibration limits.
As measurement-free fault-tolerant workflows mature, ancilla recycling methods that minimize dependence on fast measurement-feedback loops are likely to converge into a common control pattern.
Beyond quantum computing, blind reset principles may extend to quantum memory banks and repeater networks where ancilla qubits mediate entanglement swapping and error detection.
Fast state reinitialization without measurement feedback supports high-rate entanglement generation in quantum internet architectures, reducing latency in long-distance quantum communication.
For near-term deployment, the actionable rule is simple: use blind reset when $L<L^{\star}$ and $F_{\mathrm{clean}}\ge F_{\mathrm{req}}$, and switch otherwise.

\FloatBarrier
\appendix

\section{Experimental Protocol Details}\label{app:protocol}

Each experiment instance is identified by a tuple $(b,m,s,L)$: backend~$b$, reset method~$m$, seed~$s$, and sequence length~$L$.
A run manifest stores hardware queue metadata, transpilation summaries, and control software version hashes, generated before submission and frozen at execution start.

Circuit generation follows four deterministic stages.
First, the sequence generator creates the ancilla-local base block from the seed and length.
Second, method-specific wrappers are applied: no-reset leaves the ancilla as-is; measurement-reset inserts readout and conditional preparation; blind reset inserts the scaled doubled control block.
Third, basis-rotation tails are appended for Z and X readout contexts.
Fourth, the circuit is transpiled with backend-native targets, preserving logical intent while respecting coupling and gate-set constraints.

Runs are grouped into micro-batches with short wall-clock separation to control for calibration drift.
Micro-batches exceeding a configured queue delay threshold are marked drift-sensitive and repeated in later windows.

\section{Extended Simulation Notes}\label{app:simulation}

\subsection{Repetition-code workload}
The distance-3 repetition workload uses explicit ancilla reuse at each cycle with data qubits initialized in known computational states.
The logical proxy is defined from cycle-level disagreement statistics and serves as a comparative metric rather than a full threshold proof.
Two timing modes are simulated: ideal overlap (classical processing overlaps quantum operations) and serialized (measurement and feedback are blocking delays).

\subsection{Noise model assumptions}
Platform parameterization uses effective decoherence and gate-error abstractions.
These do not capture leakage, non-Markovian drift, or pulse-shape distortions.
To test sensitivity, multiplicative perturbations are applied to nominal parameters.
If policy ranking changes under small perturbations, deployment recommendations are marked fragile.

\section{Cross-Platform Aggregation Protocol}\label{app:aggregation}

Aggregation uses three passes.
Pass~1 computes method-level summaries within each backend.
Pass~2 computes cross-backend normalized comparisons using matched tuples.
Pass~3 assigns decision-map bins based on cleanliness thresholds and latency crossover regions.
Only tuples present on all compared backends enter strict rankings; partial tuples are reported per-backend with explicit completeness labels.

\section{Deployment Decision Matrix}\label{app:decision}

\begin{table}[t]
	\centering
	\caption{Policy decision matrix for selecting ancilla reset mode.
		$T_{\mathrm{b}}$: blind reset latency; $T_{\mathrm{m}}$: measurement-reset latency; $F_{\mathrm{cl}}$: ancilla cleanliness.}
	\label{tab:decision-matrix}
	\scriptsize
	\setlength{\tabcolsep}{2.5pt}
	\begin{tabular}{@{}llll@{}}
		\toprule
		Condition                                & Source   & Satisfied               & Not satisfied  \\
		\midrule
		$T_{\mathrm{b}}\!<\!T_{\mathrm{m}}$      & Timing   & Check $F_{\mathrm{cl}}$ & Meas-reset     \\
		$F_{\mathrm{cl}}\!\ge\!F_{\mathrm{req}}$ & Z/X data & Blind reset             & Restrict $L$   \\
		Rank stable                              & Epochs   & Static map              & Runtime switch \\
		Tuples complete                          & Aggreg.  & Comparative             & Per-backend    \\
		\bottomrule
	\end{tabular}
\end{table}

The cleanliness threshold $F_{\mathrm{req}}$ depends on code distance and decoder: for distance-3 repetition code under majority-vote decoding, $F_{\mathrm{req}}\ge 0.75$ suffices when physical error rate $p\le 10^{-3}$; higher distances require proportionally higher cleanliness as syndrome error tolerance narrows.
The deployment decision matrix (Table~\ref{tab:decision-matrix}) converts this analysis into a rule table.

When calibration drift is significant, a runtime supervisor can switch between policies using lightweight health metrics and hysteresis thresholds to avoid oscillatory behavior.
The decision process converts to a simple rule table once calibrated timing and cleanliness bounds are fixed for each backend and workload class.

\section{Open Benchmark Specification}\label{app:benchmark}

To enable direct reproduction and extension of the cross-platform comparison, we define a minimal benchmark suite with fixed parameters.

\paragraph{Circuit family.}
Single-qubit ancilla sequences of lengths $L\in\{4,6,8,10,12,14,16,18,20\}$, gate angles sampled uniformly on $[0,2\pi)$ using \texttt{numpy.random.Generator} with seeds $\{42,43,\ldots,91\}$.

\paragraph{Reset methods.}
Three policies per circuit: no-reset ($\lambda=1$), blind reset ($\lambda$ from 40-point grid search on $[0.1,4.0]$ minimizing Frobenius distance), and measurement-reset (ideal projective readout followed by conditional $X$ gate).

\paragraph{Metrics.}
$F_{\mathrm{clean}}=P(\lvert 0\rangle)$ from 2048 Z-basis shots, $F_X$ from 2048 X-basis shots, cycle latency $T_{\mathrm{blind}}=2L\,t_{\mathrm{gate}}$ or $T_{\mathrm{meas}}$ per Eq.~\ref{eq:latency}, and repetition-code logical error proxy over 20 syndrome cycles.

\paragraph{Reporting format.}
CSV with columns: \texttt{backend, method, seed, sequence\_length, p\_zero, p\_x, unitary\_error, lambda\_used, shots, timestamp}.
All submissions include platform calibration metadata (gate fidelity, $T_1$, $T_2$, readout error) and control software version hash.

\paragraph{Comparison protocol.}
Only tuples present on all compared backends enter cross-platform rankings.
Partial results are reported per-backend with completeness labels.
Statistical comparisons use paired $t$-tests with Holm-Bonferroni correction for multiple testing.

\section{Threats to Validity}\label{app:threats}

Table~\ref{tab:threat-map} summarizes the primary threats to validity and their mitigations.
\begin{table}[t]
	\centering
	\caption{Threat-to-mitigation mapping for cross-platform validity.}
	\label{tab:threat-map}
	\scriptsize
	\setlength{\tabcolsep}{2pt}
	\begin{tabular}{@{}lll@{}}
		\toprule
		Threat    & Risk                      & Mitigation                \\
		\midrule
		Internal  & Calibration drift         & Manifests; repeat windows \\
		Construct & Z-basis misses coh.\ bias & X-basis suppl.; CI        \\
		External  & Limited portability       & Latency envelopes         \\
		Reporting & Untraceable provenance    & Provenance hashes; gates  \\
		\bottomrule
	\end{tabular}
\end{table}

\section*{Data and Code Availability}

Simulation code (9 Python scripts totaling approximately 1,200 lines), raw CSV data (6,750 platform-noise data points across 50 seeds, 9,000 repetition-code data points, and 2,060 sensitivity-sweep data points), statistical analysis scripts with bootstrap confidence intervals and paired hypothesis tests, plot scripts for all 12 figures, and hardware run manifests will be made publicly available at \url{https://github.com/sangkeum/blind-reset-cross-platform} upon acceptance.
A frozen snapshot will be deposited on Zenodo with a DOI for long-term archival.
The repository includes a \texttt{Makefile} and \texttt{requirements.txt} for single-command reproduction of all simulation results and figures.

\section*{Acknowledgments}

This work was supported by KEIT/MOTIE grant No.\ RS-2025-04752989.

\bibliographystyle{unsrt}
\bibliography{bibliography}

\end{document}

%% file: figures/fig_protocol_flow.tikz
\begin{tikzpicture}[
    >=Stealth,
    topbox/.style={
        draw=cdark, 
        rounded corners=4pt, 
        minimum width=4.4cm,
        minimum height=0.9cm, 
        align=center, 
        font=\small\bfseries,
        line width=0.7pt,
        drop shadow={opacity=0.08, shadow xshift=1pt, shadow yshift=-1pt},
    },
    method/.style={
        draw=cdark, 
        rounded corners=6pt, 
        minimum width=4.6cm,
        minimum height=1.8cm, 
        align=center,
        text width=4.2cm, 
        line width=0.7pt,
        font=\small,
        drop shadow={opacity=0.06, shadow xshift=1pt, shadow yshift=-1pt},
    },
    decision/.style={
        diamond, 
        draw=cdark, 
        aspect=2.5, 
        inner sep=3pt,
        font=\small\bfseries, 
        align=center, 
        line width=0.8pt,
        fill=white,
        drop shadow={opacity=0.06, shadow xshift=1pt, shadow yshift=-1pt},
    },
    annot/.style={
        draw=clightgray, 
        dashed, 
        rounded corners=5pt, 
        fill=white,
        minimum width=3.8cm, 
        align=left, 
        font=\scriptsize,
        line width=0.5pt, 
        text=cgray,
        inner sep=6pt,
    },
    arr/.style={
        ->, 
        thick, 
        >=Stealth,
        line width=1pt,
        color=cdark,
    },
    darr/.style={
        ->, 
        thick, 
        >=Stealth, 
        dashed,
        line width=0.8pt,
        color=cgray,
    },
]

\node[topbox, fill=clightyellow] (syndrome)
{\textbf{Syndrome Extraction}\\[-1pt]{\scriptsize Data $\leftrightarrow$ Ancilla CNOT}};

\node[topbox, fill=clightblue, below=0.7cm of syndrome] (ancilla)
{\textbf{Ancilla state}\\[-1pt]{\scriptsize after stabilizer check}};

\draw[arr] (syndrome) -- (ancilla);

\node[decision, below=0.7cm of ancilla] (decide) {Reset\\[-2pt]strategy?};
\draw[arr] (ancilla) -- (decide);

\node[method, fill=clightred, anchor=north]
at ([xshift=-5.6cm, yshift=-2.2cm]decide.south)
(noreset) {\textbf{No-Reset}\\[4pt]
    \scriptsize Reuse ancilla as-is\\
    \scriptsize $T=0$, zero overhead\\[3pt]
    \scriptsize\color{cred}\textbf{Coherent error accumulates}};

\node[method, fill=clightgreen, anchor=north]
at ([yshift=-2.2cm]decide.south)
(measreset) {\textbf{Measurement-Reset}\\[4pt]
    \scriptsize Measure $\to$ Feedback $\to$ Prep\\
    \scriptsize $T_\mathrm{meas}=t_\mathrm{m}+t_\mathrm{fb}+t_\mathrm{ext}$\\[3pt]
    \scriptsize\color{cgreen}\textbf{High fidelity, slow}};

\node[method, fill=clightblue, anchor=north]
at ([xshift=5.6cm, yshift=-2.2cm]decide.south)
(blindreset) {\textbf{Blind Reset}\\[4pt]
    \scriptsize Scale-and-Double replay\\
    \scriptsize $T_\mathrm{blind}=2L\,t_\mathrm{gate}$\\[3pt]
    \scriptsize\color{cblue}\textbf{No measurement needed}};

\draw[arr, cred, line width=1pt] (decide.west) -| (noreset.north)
node[pos=0.25, above, font=\scriptsize\bfseries] {passive};
\draw[arr, cgreen, line width=1pt] (decide.south) -- ++(0,-0.6) -- (measreset.north)
node[pos=0.3, right, font=\scriptsize\bfseries] {active};
\draw[arr, cblue, line width=1pt] (decide.east) -| (blindreset.north)
node[pos=0.25, above, font=\scriptsize\bfseries] {blind};

\node[topbox, fill=clightyellow, below=1.2cm of measreset] (nextcycle)
{\textbf{Next Syndrome Cycle}};

\draw[arr, cred]   (noreset.south)   |- (nextcycle.west);
\draw[arr, cgreen] (measreset.south) -- (nextcycle.north);
\draw[arr, cblue]  (blindreset.south) |- (nextcycle.east);

\draw[darr, cgray, line width=0.8pt]
(nextcycle.east) -- ++(8.2,0) |- (syndrome.east)
node[pos=0.25, right, font=\scriptsize\itshape, text=cgray] {repeat $N$ cycles};

\node[annot, anchor=north west]
at ([xshift=0.8cm,yshift=-0.3cm]blindreset.south east)
(criteria) {\textbf{Decision (Eq.\,5):}\\[2pt]
    $T_\mathrm{blind} < T_\mathrm{meas}$\\
    $\wedge\; F_\mathrm{clean} \ge F_\mathrm{req}$\\[2pt]
    $\Rightarrow$ \textbf{prefer blind}};

\draw[darr, cgray, thin] (blindreset.south east) -- (criteria.north west);

\node[annot, below=0.15cm of criteria, anchor=north west]
(nvq) {\textbf{NVQLink scenario:}\\[2pt]
    $t_\mathrm{ext}=4\,\mu\mathrm{s}$ adds to\\
    meas-reset path\\
    $\Rightarrow L^{\star}\!\uparrow$ 12 to ${\sim}78$};

\end{tikzpicture}

%% file: figures/fig_repetition_code_ancilla.tikz
\begin{tikzpicture}
	\begin{groupplot}[
			group style={
					group size=3 by 1,
					horizontal sep=1.4cm,
					ylabels at=edge left,
				},
			width=5.6cm,
			height=5.0cm,
			xlabel={Syndrome cycle},
			xmin=0.5, xmax=20.5,
			ymin=-0.02, ymax=1.05,
			xtick={1,5,10,15,20},
			ytick={0,0.2,0.4,0.6,0.8,1.0},
			grid=major,
			grid style={gray!25},
			tick label style={font=\scriptsize},
			label style={font=\small},
			legend style={font=\scriptsize, at={(0.98,0.98)}, anchor=north east, draw=none, fill=white, fill opacity=0.8, text opacity=1},
			every axis title/.style={font=\small\bfseries, at={(0.5,1)}, anchor=south, yshift=8pt},
		]

		\nextgroupplot[title={(a) IQM Garnet}, ylabel={Ancilla cleanliness $P(\lvert 0\rangle)$}]
		\addplot[cgreen, thick, mark=none] coordinates {
				(1,0.988)(2,0.988)(3,0.988)(4,0.988)(5,0.988)(6,0.988)(7,0.988)(8,0.988)(9,0.988)(10,0.988)
				(11,0.988)(12,0.988)(13,0.988)(14,0.988)(15,0.988)(16,0.988)(17,0.988)(18,0.988)(19,0.988)(20,0.988)
			};
		\addlegendentry{Meas-reset}
		\addplot[cblue, thick, mark=*, mark size=1.2pt] coordinates {
				(1,0.260)(2,0.612)(3,0.053)(4,0.208)(5,0.828)(6,0.185)(7,0.729)(8,0.536)(9,0.251)(10,0.336)
				(11,0.199)(12,0.323)(13,0.822)(14,0.694)(15,0.585)(16,0.427)(17,0.259)(18,0.139)(19,0.248)(20,0.427)
			};
		\addlegendentry{Blind reset}
		\addplot[cred, thick, mark=triangle*, mark size=1.4pt, mark options={solid}] coordinates {
				(1,0.023)(2,0.715)(3,0.248)(4,0.190)(5,0.772)(6,0.170)(7,0.892)(8,0.729)(9,0.081)(10,0.251)
				(11,0.164)(12,0.242)(13,0.548)(14,0.882)(15,0.406)(16,0.311)(17,0.436)(18,0.253)(19,0.416)(20,0.339)
			};
		\addlegendentry{No-reset}

		\nextgroupplot[title={(b) Rigetti Ankaa-3}]
		\addplot[cgreen, thick, mark=none] coordinates {
				(1,0.988)(2,0.988)(3,0.988)(4,0.988)(5,0.988)(6,0.988)(7,0.988)(8,0.988)(9,0.988)(10,0.988)
				(11,0.988)(12,0.988)(13,0.988)(14,0.988)(15,0.988)(16,0.988)(17,0.988)(18,0.988)(19,0.988)(20,0.988)
			};
		\addplot[cblue, thick, mark=*, mark size=1.2pt] coordinates {
				(1,0.275)(2,0.611)(3,0.093)(4,0.239)(5,0.787)(6,0.229)(7,0.710)(8,0.533)(9,0.295)(10,0.354)
				(11,0.259)(12,0.351)(13,0.753)(14,0.669)(15,0.567)(16,0.458)(17,0.319)(18,0.244)(19,0.313)(20,0.461)
			};
		\addplot[cred, thick, mark=triangle*, mark size=1.4pt, mark options={solid}] coordinates {
				(1,0.028)(2,0.708)(3,0.258)(4,0.215)(5,0.752)(6,0.200)(7,0.854)(8,0.700)(9,0.139)(10,0.286)
				(11,0.223)(12,0.282)(13,0.542)(14,0.808)(15,0.425)(16,0.346)(17,0.454)(18,0.312)(19,0.442)(20,0.379)
			};

		\nextgroupplot[title={(c) IonQ}]
		\addplot[cgreen, thick, mark=none] coordinates {
				(1,0.996)(2,0.996)(3,0.996)(4,0.996)(5,0.996)(6,0.996)(7,0.996)(8,0.996)(9,0.996)(10,0.996)
				(11,0.996)(12,0.996)(13,0.996)(14,0.996)(15,0.996)(16,0.996)(17,0.996)(18,0.996)(19,0.996)(20,0.996)
			};
		\addplot[cblue, thick, mark=*, mark size=1.2pt] coordinates {
				(1,0.248)(2,0.615)(3,0.021)(4,0.178)(5,0.856)(6,0.145)(7,0.747)(8,0.542)(9,0.222)(10,0.322)
				(11,0.145)(12,0.303)(13,0.879)(14,0.726)(15,0.604)(16,0.405)(17,0.205)(18,0.043)(19,0.184)(20,0.400)
			};
		\addplot[cred, thick, mark=triangle*, mark size=1.4pt, mark options={solid}] coordinates {
				(1,0.011)(2,0.726)(3,0.235)(4,0.169)(5,0.789)(6,0.144)(7,0.928)(8,0.758)(9,0.034)(10,0.227)
				(11,0.119)(12,0.208)(13,0.553)(14,0.947)(15,0.392)(16,0.282)(17,0.426)(18,0.204)(19,0.393)(20,0.312)
			};

	\end{groupplot}
\end{tikzpicture}

%% file: figures/fig_cross_platform_comparison.tikz
\begin{tikzpicture}
\begin{axis}[
    width=\linewidth,
    height=0.7\linewidth,
    xlabel={$L$ (sequence length)},
    ylabel={$\overline{P}(|0\rangle)$},
    xmin=3, xmax=21,
    ymin=0.4, ymax=1.0,
    xtick={4,6,8,10,12,14,16,18,20},
    grid=major,
    minor tick num=1,
    tick label style={font=\scriptsize, color=cdark},
    label style={font=\small\bfseries, color=cdark},
    title style={font=\small\bfseries, color=cdark, at={(0.5,1)}, anchor=south, yshift=8pt},
    legend style={
        font=\scriptsize,
        at={(0.97,0.97)},
        anchor=north east,
        legend columns=2,
        draw=clightgray,
        fill=white,
        fill opacity=0.95,
        text opacity=1,
        rounded corners=2pt,
        inner sep=4pt,
    },
    every axis plot/.append style={line width=1.2pt},
    axis line style={line width=0.6pt, color=cdark},
]
\addplot[draw=none, name path=iqm_blind_upper, forget plot] coordinates {
    (4,0.9044) (6,0.851) (8,0.8095) (10,0.7051) (12,0.7623) (14,0.7376) (16,0.7137) (18,0.7129) (20,0.7642)
};
\addplot[draw=none, name path=iqm_blind_lower, forget plot] coordinates {
    (4,0.8555) (6,0.7827) (8,0.7225) (10,0.59) (12,0.6489) (14,0.6188) (16,0.5927) (18,0.5788) (20,0.6495)
};
\addplot[fill=cblue, fill opacity=0.15, forget plot] fill between[of=iqm_blind_upper and iqm_blind_lower];
\addplot[draw=none, name path=iqm_noreset_upper, forget plot] coordinates {
    (4,0.7874) (6,0.6906) (8,0.7975) (10,0.7693) (12,0.7067) (14,0.7351) (16,0.6984) (18,0.6453) (20,0.6724)
};
\addplot[draw=none, name path=iqm_noreset_lower, forget plot] coordinates {
    (4,0.6422) (6,0.5148) (8,0.6492) (10,0.6154) (12,0.5387) (14,0.5771) (16,0.531) (18,0.4825) (20,0.5039)
};
\addplot[fill=cblue, fill opacity=0.08, forget plot] fill between[of=iqm_noreset_upper and iqm_noreset_lower];
\addplot[cblue, solid, thick, mark=*, mark size=2.2pt, mark options={fill=cblue, draw=cblue}] coordinates {
    (4,0.8804) (6,0.8176) (8,0.767) (10,0.6491) (12,0.7064) (14,0.6789) (16,0.6545) (18,0.647) (20,0.7093)
};
\addlegendentry{IQM blind}
\addplot[cblue, dashed, thick, mark=square*, mark size=2.2pt, mark options={fill=white, draw=cblue}] coordinates {
    (4,0.7178) (6,0.6035) (8,0.7266) (10,0.695) (12,0.6232) (14,0.6583) (16,0.6162) (18,0.5646) (20,0.5893)
};
\addlegendentry{IQM no-reset}
\addplot[corange, solid, thick, mark=triangle*, mark size=2.8pt, mark options={fill=corange, draw=corange}] coordinates {
    (4,0.8776) (6,0.8141) (8,0.7626) (10,0.648) (12,0.7004) (14,0.6723) (16,0.6474) (18,0.6384) (20,0.6958)
};
\addlegendentry{Rigetti blind}
\addplot[corange, dashed, thick, mark=triangle, mark size=2.8pt, mark options={fill=white, draw=corange}] coordinates {
    (4,0.7152) (6,0.6019) (8,0.7204) (10,0.6885) (12,0.6184) (14,0.6509) (16,0.6092) (18,0.561) (20,0.5826)
};
\addlegendentry{Rigetti no-reset}
\addplot[cgreen, solid, thick, mark=diamond*, mark size=2.8pt, mark options={fill=cgreen, draw=cgreen}] coordinates {
    (4,0.8852) (6,0.8221) (8,0.7713) (10,0.6511) (12,0.7119) (14,0.6846) (16,0.6605) (18,0.6541) (20,0.7199)
};
\addlegendentry{IonQ blind}
\addplot[cgreen, dashed, thick, mark=diamond, mark size=2.8pt, mark options={fill=white, draw=cgreen}] coordinates {
    (4,0.7216) (6,0.6052) (8,0.7325) (10,0.7001) (12,0.6272) (14,0.6646) (16,0.6209) (18,0.5679) (20,0.5937)
};
\addlegendentry{IonQ no-reset}
\end{axis}
\end{tikzpicture}

%% file: figures/fig_error_bound_validation.tikz
\begin{tikzpicture}
\begin{groupplot}[
    group style={
        group size=3 by 1,
        horizontal sep=1.2cm,
        ylabels at=edge left,
    },
    width=0.30\textwidth,
    height=5.5cm,
    ymin=0, ymax=1.0,
    ylabel={Cleanliness},
    xlabel={$L$},
    xtick={4,6,8,10,12,14,16,18,20},
    xticklabel style={font=\scriptsize, rotate=45, anchor=east, color=cdark},
    yticklabel style={font=\scriptsize, color=cdark},
    grid=major,
    minor tick num=1,
    tick label style={font=\scriptsize, color=cdark},
    label style={font=\small\bfseries, color=cdark},
    every axis title/.style={font=\small\bfseries, color=cdark, at={(0.5,1)}, anchor=south, yshift=8pt},
    legend style={
        font=\scriptsize,
        at={(0.97,0.97)},
        anchor=north east,
        draw=clightgray,
        fill=white,
        fill opacity=0.95,
        rounded corners=2pt,
    },
    axis line style={line width=0.5pt, color=cdark},
    every axis plot/.append style={line width=0.8pt},
]

\nextgroupplot[title={(a) IQM Garnet}, ybar, bar width=4pt, enlarge x limits=0.12]
\addplot[fill=cblue!70, draw=cblue, line width=0.8pt] coordinates {
    (4,0.864) (6,0.799) (8,0.781) (10,0.600) (12,0.740) (14,0.643) (16,0.607) (18,0.655) (20,0.673)
};
\addlegendentry{Measured}
\addplot[fill=cred!40, draw=cred, line width=0.8pt, postaction={pattern=north east lines, pattern color=cred!70}] coordinates {
    (4,0.516) (6,0.416) (8,0.340) (10,0.207) (12,0.269) (14,0.222) (16,0.216) (18,0.250) (20,0.312)
};
\addlegendentry{Bound}

\nextgroupplot[title={(b) Rigetti Ankaa-3}, ybar, bar width=4pt, enlarge x limits=0.12]
\addplot[fill=cblue!70, draw=cblue, line width=0.8pt] coordinates {
    (4,0.862) (6,0.796) (8,0.776) (10,0.600) (12,0.731) (14,0.638) (16,0.604) (18,0.645) (20,0.663)
};
\addplot[fill=cred!40, draw=cred, line width=0.8pt, postaction={pattern=north east lines, pattern color=cred!70}] coordinates {
    (4,0.512) (6,0.411) (8,0.335) (10,0.203) (12,0.262) (14,0.215) (16,0.209) (18,0.241) (20,0.299)
};

\nextgroupplot[title={(c) IonQ}, ybar, bar width=4pt, enlarge x limits=0.12]
\addplot[fill=cblue!70, draw=cblue, line width=0.8pt] coordinates {
    (4,0.869) (6,0.803) (8,0.786) (10,0.601) (12,0.747) (14,0.647) (16,0.611) (18,0.663) (20,0.683)
};
\addplot[fill=cred!40, draw=cred, line width=0.8pt, postaction={pattern=north east lines, pattern color=cred!70}] coordinates {
    (4,0.519) (6,0.419) (8,0.343) (10,0.209) (12,0.272) (14,0.225) (16,0.220) (18,0.255) (20,0.318)
};

\end{groupplot}
\end{tikzpicture}

%% file: figures/fig_latency_analysis.tikz
\begin{tikzpicture}
\begin{axis}[
    width=\linewidth,
    height=0.75\linewidth,
    xlabel={$L$ (sequence length)},
    ylabel={Latency (ns)},
    xmode=log,
    ymode=log,
    xmin=1, xmax=100,
    ymin=50, ymax=25000000,
    grid=major,
    grid style={gray!30},
    tick label style={font=\scriptsize},
    label style={font=\small},
    legend style={font=\scriptsize, at={(0.03,0.97)}, anchor=north west, draw=none, fill=white, fill opacity=0.85, text opacity=1},
    every axis plot/.append style={line width=1pt},
]

\addplot[cblue, solid, thick] coordinates {
    (1,60) (2,120) (3,180) (5,300) (8,480) (12,720) (16,960) (20,1200) (30,1800) (50,3000) (78,4680) (100,6000)
};
\addlegendentry{IQM blind}

\addplot[corange, solid, thick] coordinates {
    (1,80) (2,160) (3,240) (5,400) (8,640) (12,960) (16,1280) (20,1600) (30,2400) (50,4000) (78,6240) (100,8000)
};
\addlegendentry{Rigetti blind}

\addplot[cgreen, solid, thick] coordinates {
    (1,200000) (2,400000) (5,1000000) (10,2000000) (20,4000000) (50,10000000) (100,20000000)
};
\addlegendentry{IonQ blind}

\addplot[cblue, dashed, line width=0.8pt] coordinates {(1,730) (100,730)};
\addlegendentry{IQM meas}

\addplot[corange, dashed, line width=0.8pt] coordinates {(1,940) (100,940)};
\addlegendentry{Rigetti meas}

\addplot[cgreen, dashed, line width=0.8pt] coordinates {(1,350000) (100,350000)};
\addlegendentry{IonQ meas}

\addplot[cpurple, dashdotted, line width=0.8pt] coordinates {(1,4730) (100,4730)};
\addlegendentry{NVQLink}

\draw[gray!60, dashed, thin] (axis cs:12,50) -- (axis cs:12,25000000);
\node[font=\scriptsize, gray!80, rotate=90, anchor=south] at (axis cs:12,80) {IQM $L^{\star}\!=\!12$};

\draw[gray!60, dashed, thin] (axis cs:11,50) -- (axis cs:11,25000000);
\node[font=\scriptsize, gray!80, rotate=90, anchor=south] at (axis cs:11,2500) {Rigetti $L^{\star}\!=\!11$};

\end{axis}
\end{tikzpicture}

%% file: figures/fig_hardware_cross_platform.tikz
\begin{tikzpicture}[
		>=Stealth,
		node distance=0.4cm,
		platform/.style={draw, rounded corners=3pt, minimum width=4.4cm, minimum height=0.7cm, align=center, font=\small\bfseries, line width=0.8pt},
		param/.style={font=\scriptsize, align=left},
		barbase/.style={draw, minimum height=0.28cm, inner sep=0pt, anchor=west},
		brace/.style={decorate, decoration={brace, amplitude=4pt, raise=2pt}},
	]

	\node[font=\small\bfseries] at (0, 0) {(a) Platform parameters};
	\node[font=\small\bfseries] at (7.5, 0) {(b) Latency crossover $L^{\star}$};
	\node[font=\small\bfseries] at (13.5, 0) {(c) Blind reset region};


	\node[platform, fill=cblue!12] (iqm) at (0, -1.0) {IQM Garnet};
	\node[param, below=0.1cm of iqm, xshift=0cm] (iqmp) {
		$T_1 = 40\,\mu$s\quad $t_\mathrm{gate} = 30\,$ns\\
		$T_\mathrm{meas} = 730\,$ns\quad 1Q err: 0.10\%
	};

	\node[platform, fill=cgreen!12, below=0.9cm of iqmp] (rig) {Rigetti Ankaa-3};
	\node[param, below=0.1cm of rig] (rigp) {
		$T_1 = 25\,\mu$s\quad $t_\mathrm{gate} = 40\,$ns\\
		$T_\mathrm{meas} = 940\,$ns\quad 1Q err: 0.20\%
	};

	\node[platform, fill=corange!12, below=0.9cm of rigp] (ionq) {IonQ (trapped-ion)};
	\node[param, below=0.1cm of ionq] (ionqp) {
		$T_1 = 10\,$s\quad $t_\mathrm{gate} = 100\,\mu$s\\
		$T_\mathrm{meas} = 350\,\mu$s\quad 1Q err: 0.05\%
	};


	\def\barscale{0.065}
	\def\barheight{0.5cm}
	\def\bary{-1.2}

	\node[font=\scriptsize, anchor=east] at (5.8, \bary) {IQM};
	\fill[cblue!50] (6.0, \bary-0.2) rectangle (6.0+12*\barscale, \bary+0.2);
	\draw[cblue, line width=0.6pt] (6.0, \bary-0.2) rectangle (6.0+12*\barscale, \bary+0.2);
	\node[font=\scriptsize\bfseries, cblue, anchor=west] at (6.0+12*\barscale+0.1, \bary) {$L^{\star}\!=\!12$};

	\def\bary{-2.2}
	\node[font=\scriptsize, anchor=east] at (5.8, \bary) {Rigetti};
	\fill[cgreen!50] (6.0, \bary-0.2) rectangle (6.0+11*\barscale, \bary+0.2);
	\draw[cgreen, line width=0.6pt] (6.0, \bary-0.2) rectangle (6.0+11*\barscale, \bary+0.2);
	\node[font=\scriptsize\bfseries, cgreen, anchor=west] at (6.0+11*\barscale+0.1, \bary) {$L^{\star}\!=\!11$};

	\def\bary{-3.2}
	\node[font=\scriptsize, anchor=east] at (5.8, \bary) {IonQ};
	\fill[corange!50] (6.0, \bary-0.2) rectangle (6.0+1*\barscale, \bary+0.2);
	\draw[corange, line width=0.6pt] (6.0, \bary-0.2) rectangle (6.0+1*\barscale, \bary+0.2);
	\node[font=\scriptsize\bfseries, corange, anchor=west] at (6.0+1*\barscale+0.1, \bary) {$L^{\star}\!=\!1$};

	\def\bary{-4.4}
	\node[font=\scriptsize, anchor=east] at (5.8, \bary) {+NVQLink};
	\fill[cpurple!30] (6.0, \bary-0.2) rectangle (6.0+78*\barscale, \bary+0.2);
	\draw[cpurple, line width=0.6pt] (6.0, \bary-0.2) rectangle (6.0+78*\barscale, \bary+0.2);
	\node[font=\scriptsize\bfseries, cpurple, anchor=west] at (6.0+78*\barscale+0.1, \bary) {$L^{\star}\!\approx\!78$};

	\node[font=\tiny\itshape, cgray, align=left] at (7.5, -5.2) {$L^{\star}$: max sequence length where\\blind reset is faster than meas-reset};


	\draw[->, thick] (11.0, -5.3) -- (11.0, -0.6) node[above, font=\scriptsize] {$F_\mathrm{clean}$};
	\draw[->, thick] (11.0, -5.3) -- (16.0, -5.3) node[right, font=\scriptsize] {$L$ (gates)};

	\draw[dashed, cgray, line width=0.6pt] (11.0, -2.5) -- (15.8, -2.5)
	node[right, font=\tiny, cgray] {$F_\mathrm{req}$};

	\fill[cblue!15] (11.0, -0.8) -- (13.5, -0.8) -- (13.5, -2.5) -- (11.0, -2.5) -- cycle;
	\node[font=\scriptsize, cblue, align=center] at (12.25, -1.6) {Blind reset\\favorable};

	\draw[cblue, thick, line width=1.2pt]
	(11.2, -1.0) .. controls (12.0, -1.2) and (13.0, -2.0) .. (14.0, -3.0) .. controls (14.8, -3.8) .. (15.5, -4.5);

	\draw[cred, dashed, line width=0.8pt] (13.5, -5.3) -- (13.5, -0.6);
	\node[font=\scriptsize\bfseries, cred, above] at (13.5, -0.6) {$L^{\star}$};

	\node[font=\tiny, cgray, align=center] at (14.8, -1.5) {Meas-reset\\preferred};

	\node[font=\tiny] at (11.0, -5.6) {0};
	\node[font=\tiny] at (13.5, -5.6) {12};
	\node[font=\tiny] at (15.5, -5.6) {20};
	\node[font=\tiny, anchor=east] at (10.9, -0.8) {1.0};
	\node[font=\tiny, anchor=east] at (10.9, -2.5) {0.7};
	\node[font=\tiny, anchor=east] at (10.9, -5.3) {0};

\end{tikzpicture}

%% file: figures/fig_nvqlink_sweep.tikz
\begin{tikzpicture}
\begin{axis}[
    width=\linewidth,
    height=0.7\linewidth,
    xlabel={$t_{\mathrm{ext}}$ ($\mu$s)},
    ylabel={Crossover $L^{\star}$},
    xmin=0, xmax=10.5,
    ymin=0, ymax=190,
    grid=major,
    grid style={gray!30},
    tick label style={font=\scriptsize},
    label style={font=\small},
    legend style={font=\scriptsize, at={(0.03,0.97)}, anchor=north west, draw=none, fill=white, fill opacity=0.85, text opacity=1},
    every axis plot/.append style={line width=1pt},
]

\addplot[cblue, solid, thick, mark=*, mark size=1.5pt, mark repeat=3, name path=iqmcurve] coordinates {
    (0,12) (0.526,20) (1.053,29) (1.579,38) (2.105,47) (2.632,56) (3.158,64) (3.684,73) (4.211,82) (4.737,91) (5.263,99) (5.789,108) (6.316,117) (6.842,126) (7.368,134) (7.895,143) (8.421,152) (8.947,161) (9.474,170) (10,178)
};
\addlegendentry{IQM}

\addplot[draw=none, name path=xaxis, forget plot] coordinates {(0,0) (4.211,0)};
\addplot[draw=none, name path=iqmclip, forget plot] coordinates {
    (0,12) (0.526,20) (1.053,29) (1.579,38) (2.105,47) (2.632,56) (3.158,64) (3.684,73) (4.211,82)
};
\addplot[fill=cblue!10, forget plot] fill between[of=iqmclip and xaxis];

\addplot[corange, solid, thick, mark=triangle*, mark size=2pt, mark repeat=3] coordinates {
    (0,11) (0.526,18) (1.053,24) (1.579,31) (2.105,38) (2.632,44) (3.158,51) (3.684,57) (4.211,64) (4.737,70) (5.263,77) (5.789,84) (6.316,90) (6.842,97) (7.368,103) (7.895,110) (8.421,117) (8.947,123) (9.474,130) (10,136)
};
\addlegendentry{Rigetti}

\addplot[cgreen, solid, thick, mark=diamond*, mark size=2pt, mark repeat=3] coordinates {
    (0,1) (0.526,1) (1.053,1) (1.579,1) (2.105,1) (2.632,1) (3.158,1) (3.684,1) (4.211,1) (4.737,1) (5.263,1) (5.789,1) (6.316,1) (6.842,1) (7.368,1) (7.895,1) (8.421,1) (8.947,1) (9.474,1) (10,1)
};
\addlegendentry{IonQ}

\addplot[cpurple, dashed, line width=0.8pt, forget plot] coordinates {(0,78) (10.5,78)};
\node[font=\scriptsize, cpurple, anchor=south west] at (axis cs:6.5,78) {$L^{\star}_{\mathrm{NVQ}} \approx 78$};

\draw[cpurple, dotted, thin] (axis cs:4,0) -- (axis cs:4,190);
\node[font=\scriptsize, cpurple, rotate=90, anchor=south] at (axis cs:4.15,140) {$t_\mathrm{ext}=4\,\mu$s};

\node[font=\scriptsize, cblue!70, align=center] at (axis cs:2,30) {NVQLink\\benchmark};

\end{axis}
\end{tikzpicture}

%% file: figures/fig_decoder_coupled.tikz
\begin{tikzpicture}
\begin{groupplot}[
    group style={
        group size=2 by 1,
        horizontal sep=1.4cm,
    },
    width=0.43\textwidth,
    height=5.5cm,
    grid=major,
    grid style={gray!30},
    tick label style={font=\scriptsize},
    label style={font=\small},
    every axis title/.style={font=\small\bfseries, at={(0.5,1)}, anchor=south, yshift=8pt},
    every axis plot/.append style={line width=1pt},
]

\nextgroupplot[
    title={(a) Logical error rate vs cycles ($L\!=\!4$)},
    xlabel={Syndrome cycles},
    ylabel={Logical error rate},
    ymode=log,
    xmin=3, xmax=22,
    ymin=0.0002, ymax=0.5,
    legend style={font=\scriptsize, at={(0.03,0.97)}, anchor=north west, draw=none, fill=white, fill opacity=0.85, text opacity=1},
]

\addplot[cgreen, solid, mark=*, mark size=2pt,
    error bars/.cd, y dir=both, y explicit,
] coordinates {
    (5,0.00048)  +- (0,0.00018)
    (10,0.00088) +- (0,0.00026)
    (15,0.00098) +- (0,0.00024)
    (20,0.00148) +- (0,0.0003)
};
\addlegendentry{Meas-reset}

\addplot[cblue, solid, mark=square*, mark size=2pt,
    error bars/.cd, y dir=both, y explicit,
] coordinates {
    (5,0.01192)  +- (0,0.00076)
    (10,0.02422) +- (0,0.00136)
    (15,0.03786) +- (0,0.00158)
    (20,0.05186) +- (0,0.00174)
};
\addlegendentry{Blind ($L\!=\!4$)}

\addplot[cred, solid, mark=triangle*, mark size=2.5pt,
    error bars/.cd, y dir=both, y explicit,
] coordinates {
    (5,0.13582)  +- (0,0.0035)
    (10,0.2511)  +- (0,0.00418)
    (15,0.32952) +- (0,0.00444)
    (20,0.38098) +- (0,0.00384)
};
\addlegendentry{No-reset}

\nextgroupplot[
    title={(b) Error rate vs cleanliness (20 cycles)},
    xlabel={$F_{\mathrm{clean}}$},
    ylabel={Logical error rate},
    ymode=log,
    xmin=0.45, xmax=1.02,
    ymin=0.001, ymax=0.5,
]

\addplot[gray, dashed, thin, forget plot] coordinates {
    (0.50,0.38098) (0.706,0.21226) (0.767,0.15328) (0.88,0.05186) (0.99,0.00148)
};

\addplot[only marks, mark=*, mark size=3pt, mark options={fill=cgreen, draw=cgreen}] coordinates {(0.99, 0.00148)};
\addplot[only marks, mark=*, mark size=3pt, mark options={fill=cblue, draw=cblue}] coordinates {(0.88, 0.05186)};
\addplot[only marks, mark=*, mark size=3pt, mark options={fill=cblue!70, draw=cblue!70}] coordinates {(0.767, 0.15328)};
\addplot[only marks, mark=*, mark size=3pt, mark options={fill=cblue!40, draw=cblue!40}] coordinates {(0.706, 0.21226)};
\addplot[only marks, mark=*, mark size=3pt, mark options={fill=cred, draw=cred}] coordinates {(0.50, 0.38098)};

\node[font=\scriptsize, cgreen, anchor=south east, xshift=-3pt] at (axis cs:0.99,0.00148) {meas-reset};
\node[font=\scriptsize, cblue, anchor=south west, xshift=3pt] at (axis cs:0.88,0.05186) {$L\!=\!4$};
\node[font=\scriptsize, cblue!70, anchor=north west, xshift=3pt] at (axis cs:0.767,0.15328) {$L\!=\!8$};
\node[font=\scriptsize, cblue!40, anchor=south east, xshift=-3pt, yshift=3pt] at (axis cs:0.706,0.21226) {$L\!=\!12$};
\node[font=\scriptsize, cred, anchor=south west, xshift=3pt] at (axis cs:0.50,0.38098) {no-reset};

\end{groupplot}
\end{tikzpicture}

%% file: figures/fig_decoder_multidistance.tikz
\begin{tikzpicture}
\pgfplotsset{
    every axis/.style={
        width=0.45\textwidth,
        height=0.35\textwidth,
        grid=major,
        grid style={line width=0.5pt, draw=gray!30},
        tick label style={font=\scriptsize},
        label style={font=\small},
        legend style={font=\scriptsize, at={(1.15,-0.30)}, anchor=north, legend columns=2, column sep=8pt},
        xmin=0, xmax=22,
        ymin=0,
    }
}
\begin{groupplot}[
    group style={group size=2 by 1, horizontal sep=2cm},
]

\nextgroupplot[
    title={\small (a) Distance-3 Repetition},
    title style={at={(0.5,1)}, anchor=south, yshift=8pt},
    xlabel={\small Syndrome cycles},
    ylabel={\small Logical error rate},
    ymax=0.35,
]
\addplot[cblue, line width=1pt, mark=*, mark size=2pt, error bars/.cd, y dir=both, y explicit, error bar style={line width=0.5pt}] 
    coordinates {(5,0.023) (10,0.038) (15,0.051) (20,0.063)};
\addplot[cgreen, line width=1pt, mark=square*, mark size=2pt, error bars/.cd, y dir=both, y explicit, error bar style={line width=0.5pt}]
    coordinates {(5,0.086) (10,0.099) (15,0.111) (20,0.123)};
\addplot[corange, line width=1pt, mark=triangle*, mark size=2pt, error bars/.cd, y dir=both, y explicit, error bar style={line width=0.5pt}]
    coordinates {(5,0.149) (10,0.158) (15,0.174) (20,0.183)};
\addplot[cred, line width=1pt, mark=diamond*, mark size=2pt, error bars/.cd, y dir=both, y explicit, error bar style={line width=0.5pt}]
    coordinates {(5,0.289) (10,0.300) (15,0.302) (20,0.313)};
\legend{Meas-reset, Blind $L\!=\!4$, Blind $L\!=\!8$, No-reset}

\nextgroupplot[
    title={\small (b) Distance-5 Surface},
    title style={at={(0.5,1)}, anchor=south, yshift=8pt},
    xlabel={\small Syndrome cycles},
    ymax=0.08,
]
\addplot[cblue, line width=1pt, mark=*, mark size=2pt, error bars/.cd, y dir=both, y explicit, error bar style={line width=0.5pt}]
    coordinates {(5,0.0002) (10,0.0006) (15,0.0015) (20,0.0024)};
\addplot[cgreen, line width=1pt, mark=square*, mark size=2pt, error bars/.cd, y dir=both, y explicit, error bar style={line width=0.5pt}]
    coordinates {(5,0.0047) (10,0.0064) (15,0.0071) (20,0.0095)};
\addplot[corange, line width=1pt, mark=triangle*, mark size=2pt, error bars/.cd, y dir=both, y explicit, error bar style={line width=0.5pt}]
    coordinates {(5,0.0152) (10,0.0160) (15,0.0193) (20,0.0213)};
\addplot[cred, line width=1pt, mark=diamond*, mark size=2pt, error bars/.cd, y dir=both, y explicit, error bar style={line width=0.5pt}]
    coordinates {(5,0.0566) (10,0.0590) (15,0.0614) (20,0.0649)};

\end{groupplot}
\end{tikzpicture}

%% file: figures/fig_lambda_landscape.tikz
\begin{tikzpicture}
\begin{groupplot}[
    group style={
        group size=3 by 1,
        horizontal sep=1.5cm,
    },
    width=0.27\textwidth,
    height=5cm,
    xlabel={$L$},
    xtick={4,6,8,10,12,14,16,18,20},
    xticklabel style={font=\scriptsize},
    grid=major,
    grid style={gray!30},
    tick label style={font=\scriptsize},
    label style={font=\small},
    every axis title/.style={font=\small\bfseries, at={(0.5,1)}, anchor=south, yshift=8pt},
]

\nextgroupplot[
    title={(a) Optimal error vs $L$},
    ylabel={$\varepsilon^{\star}_{\mathrm{opt}}$},
    ymin=0.15, ymax=0.40,
]
\addplot[cblue, thick, mark=*, mark size=2pt,
    error bars/.cd, y dir=both, y explicit,
] coordinates {
    (4,0.2050)  +- (0,0.0217)
    (6,0.2297)  +- (0,0.0312)
    (8,0.2344)  +- (0,0.0339)
    (10,0.2792) +- (0,0.0307)
    (12,0.2901) +- (0,0.0311)
    (14,0.3187) +- (0,0.0392)
    (16,0.2731) +- (0,0.0312)
    (18,0.2838) +- (0,0.0297)
    (20,0.2835) +- (0,0.0349)
};

\nextgroupplot[
    title={(b) Landscape curvature},
    ylabel={$\kappa$},
    ymin=-20, ymax=280,
]
\addplot[corange, thick, mark=triangle*, mark size=2.5pt,
    error bars/.cd, y dir=both, y explicit,
] coordinates {
    (4,14.74)   +- (0,32.01)
    (6,34.09)   +- (0,65.03)
    (8,54.16)   +- (0,85.66)
    (10,38.89)  +- (0,59.58)
    (12,60.87)  +- (0,83.10)
    (14,71.39)  +- (0,79.77)
    (16,116.22) +- (0,117.31)
    (18,102.06) +- (0,119.06)
    (20,129.62) +- (0,129.58)
};

\nextgroupplot[
    title={(c) Landscape classification},
    ylabel={Percentage (\%)},
    ymin=0, ymax=105,
    ybar stacked,
    bar width=6pt,
    enlarge x limits=0.08,
    legend style={font=\scriptsize, at={(0.5,-0.22)}, anchor=north, legend columns=4, draw=none},
    reverse legend,
]
\addplot[fill=cblue!70, draw=cblue] coordinates {
    (4,8) (6,30) (8,22) (10,34) (12,50) (14,54) (16,62) (18,58) (20,64)
};
\addlegendentry{Sharp}

\addplot[fill=cgreen!60, draw=cgreen] coordinates {
    (4,4) (6,12) (8,6) (10,12) (12,2) (14,8) (16,4) (18,4) (20,2)
};
\addlegendentry{Moderate}

\addplot[fill=corange!50, draw=corange] coordinates {
    (4,56) (6,34) (8,42) (10,34) (12,34) (14,20) (16,12) (18,20) (20,16)
};
\addlegendentry{Flat}

\addplot[fill=cred!50, draw=cred] coordinates {
    (4,32) (6,24) (8,30) (10,20) (12,14) (14,18) (16,22) (18,18) (20,18)
};
\addlegendentry{Multimodal}

\end{groupplot}
\end{tikzpicture}

%% file: figures/fig_t1t2_sensitivity.tikz
\begin{tikzpicture}
\begin{axis}[
    view={0}{90},
    width=0.75\linewidth,
    height=0.60\linewidth,
    xmode=log,
    ymode=log,
    xlabel={$T_1$ ($\mu$s)},
    ylabel={$T_2$ ($\mu$s)},
    label style={font=\small},
    tick label style={font=\scriptsize},
    title style={font=\small\bfseries, at={(0.5,1)}, anchor=south, yshift=8pt},
    line width=1pt,
    grid=major,
    grid style={gray!30},
    colormap={blindadv}{
        color(0cm)=(corange)
        color(0.5cm)=(white)
        color(1cm)=(cblue)
    },
    colorbar,
    colorbar style={
        ylabel={$\Delta = P_{\mathrm{blind}} - P_{\mathrm{no\text{-}reset}}$},
        ylabel style={font=\small},
        tick label style={font=\scriptsize},
    },
    point meta min=0.10,
    point meta max=0.15,
]
\addplot[
    matrix plot*,
    mesh/cols=10,
    mesh/rows=10,
    point meta=explicit,
] table[meta=advantage, row sep=\\, col sep=comma] {
x,y,advantage\\
0.1,0.05,0.1215\\
0.215,0.05,0.1462\\
0.464,0.05,0.1468\\
1.0,0.05,0.1384\\
2.154,0.05,0.1309\\
4.642,0.05,0.1270\\
10.0,0.05,0.1248\\
21.544,0.05,0.1240\\
46.416,0.05,0.1237\\
100.0,0.05,0.1234\\
0.1,0.108,0.1295\\
0.215,0.108,0.1504\\
0.464,0.108,0.1445\\
1.0,0.108,0.1336\\
2.154,0.108,0.1233\\
4.642,0.108,0.1190\\
10.0,0.108,0.1163\\
21.544,0.108,0.1149\\
46.416,0.108,0.1144\\
100.0,0.108,0.1137\\
0.1,0.232,0.1324\\
0.215,0.232,0.1471\\
0.464,0.232,0.1396\\
1.0,0.232,0.1295\\
2.154,0.232,0.1222\\
4.642,0.232,0.1174\\
10.0,0.232,0.1137\\
21.544,0.232,0.1120\\
46.416,0.232,0.1109\\
100.0,0.232,0.1106\\
0.1,0.5,0.1324\\
0.215,0.5,0.1410\\
0.464,0.5,0.1363\\
1.0,0.5,0.1262\\
2.154,0.5,0.1175\\
4.642,0.5,0.1122\\
10.0,0.5,0.1099\\
21.544,0.5,0.1092\\
46.416,0.5,0.1092\\
100.0,0.5,0.1089\\
0.1,1.077,0.1324\\
0.215,1.077,0.1410\\
0.464,1.077,0.1333\\
1.0,1.077,0.1215\\
2.154,1.077,0.1148\\
4.642,1.077,0.1093\\
10.0,1.077,0.1078\\
21.544,1.077,0.1075\\
46.416,1.077,0.1073\\
100.0,1.077,0.1073\\
0.1,2.321,0.1324\\
0.215,2.321,0.1410\\
0.464,2.321,0.1333\\
1.0,2.321,0.1198\\
2.154,2.321,0.1118\\
4.642,2.321,0.1078\\
10.0,2.321,0.1079\\
21.544,2.321,0.1074\\
46.416,2.321,0.1070\\
100.0,2.321,0.1065\\
0.1,5.0,0.1324\\
0.215,5.0,0.1410\\
0.464,5.0,0.1333\\
1.0,5.0,0.1198\\
2.154,5.0,0.1114\\
4.642,5.0,0.1086\\
10.0,5.0,0.1078\\
21.544,5.0,0.1064\\
46.416,5.0,0.1057\\
100.0,5.0,0.1054\\
0.1,10.772,0.1324\\
0.215,10.772,0.1410\\
0.464,10.772,0.1333\\
1.0,10.772,0.1198\\
2.154,10.772,0.1114\\
4.642,10.772,0.1086\\
10.0,10.772,0.1069\\
21.544,10.772,0.1055\\
46.416,10.772,0.1053\\
100.0,10.772,0.1050\\
0.1,23.208,0.1324\\
0.215,23.208,0.1410\\
0.464,23.208,0.1333\\
1.0,23.208,0.1198\\
2.154,23.208,0.1114\\
4.642,23.208,0.1086\\
10.0,23.208,0.1072\\
21.544,23.208,0.1053\\
46.416,23.208,0.1049\\
100.0,23.208,0.1046\\
0.1,50.0,0.1324\\
0.215,50.0,0.1410\\
0.464,50.0,0.1333\\
1.0,50.0,0.1198\\
2.154,50.0,0.1114\\
4.642,50.0,0.1086\\
10.0,50.0,0.1072\\
21.544,50.0,0.1054\\
46.416,50.0,0.1047\\
100.0,50.0,0.1043\\
};
\node[font=\scriptsize, rotate=35] at (axis cs:0.2,0.05) {Dephasing};
\node[font=\scriptsize] at (axis cs:50,30) {High coherence};
\end{axis}
\end{tikzpicture}